\documentclass[11pt, oneside]{article}   	
\usepackage{geometry}                		
\usepackage{graphicx}				
\usepackage{amssymb}

\usepackage{graphicx}
\usepackage{newtxtext}
\usepackage{newtxmath}
\usepackage{natbib}
\usepackage{hyperref}
\usepackage{bm}

\usepackage{algorithm}
\usepackage[noend]{algpseudocode}
\usepackage{xcolor}
\usepackage{enumerate}
\usepackage{enumitem}
\usepackage{booktabs} 

\usepackage{mathtools}
\usepackage{float}

\newcommand{\RomanNumeralCaps}[1]

\newcommand{\mtxt}[1]{\ensuremath{\mathrm{#1}}}
\newcommand*\rfrac[2]{{}^{#1}\!/_{#2}}

\newcommand{\vect}[1]{\boldsymbol{\mathbf{#1}}}
\newcommand{\E}[1]{\ensuremath{\cdot 10^{#1}}}
\newcommand{\overbar}[1]{\mkern1.5mu\overline{\mkern-2mu#1\mkern-1mu}\mkern 1.3mu}

\title{Real--time thermoacoustic data assimilation}
\author{Andrea Nóvoa\footnote{Cambridge University Engineering Department, Trumpington St, Cambridge CB2 1PZ, UK} \& Luca Magri$^*$\footnote{Imperial College London, Aeronautics Department,  
Exhibition Road, London SW7 2AZ, UK} $\,$\footnote{
The Alan Turing Institute, 96 Euston Rd, London NW1 2DB, UK}
 $\,$\footnote{Institute for Advanced Study, TU Munich, Lichtenbergstraße 2a
85748 Garching, Germany (visiting)} $\,$\footnote{lm547@cam.ac.uk} }

\begin{document}
\maketitle

\begin{abstract}
Low--order thermoacoustic models are qualitatively correct, but they are typically quantitatively inaccurate. We propose a time--domain {bias--aware} method to make qualitatively low--order models  quantitatively (more) accurate.  
First, we develop a Bayesian ensemble data assimilation method for a low--order model to self--adapt and self--correct any time that reference data becomes available. 
Second, 
we apply the methodology to infer the thermoacoustic states and heat release parameters on the fly without storing data (real--time). We perform twin experiments using synthetic acoustic pressure measurements to analyse the performance of data assimilation in all nonlinear thermoacoustic regimes, from limit cycles to chaos, and interpret the results physically. 
Third, we propose practical rules for thermoacoustic data assimilation. An {\it increase, reject, inflate} strategy is proposed to deal with the rich nonlinear behaviour; and physical time scales for assimilation are proposed in non--chaotic regimes (with the Nyquist--Shannon criterion) and in chaotic regimes (with the Lyapunov time). 
{
Fourth, we perform data assimilation using data from a higher--fidelity model. We introduce an echo state network to estimate in real--time the forecast bias, which is the model error of the low--fidelity model. }
We show that 
{
(i) the correct acoustic pressure, parameters, and model bias can be accurately inferred; } 
(ii) the learning is robust as it can tackle large uncertainties in the observations (up to 50\% the mean values);   
(iii) the uncertainty of the prediction and parameters is naturally part of the output; and 
(iv) both the time--accurate solution and statistics can be successfully inferred. 
Data assimilation opens up new possibility for real--time prediction of thermoacoustics by synergistically combining physical knowledge and experimental data. 

\end{abstract}

\section{Introduction}

When the heat released by a heat source, such as a flame, is sufficiently in phase with the acoustic waves of a confined environment, such a gas turbine or a rocket, 
 thermoacoustic oscillations may occur~\citep{rayleigh_explanation_1878}. 
Thermoacoustic oscillations manifest themselves as large--amplitude vibrations, which can be detrimental to gas--turbine reliability~\citep[e.g.,][]{lieuwen_book_2012}, and can be destructive in high--power--density motors such as rocket engines~\citep[e.g.,][]{culick_unsteady_2006}.
The objective of manufacturers is to design devices that are thermoacoustically stable, which is the goal of optimisation, and suppress a thermoacoustic oscillation if it occurs, which is the goal of control~\citep[e.g.,][]{magri_adjoint_2019}. 
Both optimisation and control rely on a mathematical model, which provides predictions on the key physical variables, such as the acoustic pressure and the heat release rate. 
The accurate prediction of thermoacoustic oscillations, however, remains one of the most challenging problems faced by power generation, heating and propulsion manufacturers~\citep[e.g.,][]{dowling_feedback_2005,noiray2008unified,lieuwen_book_2012,poinsot_prediction_2017,juniper_sensitivity_2018}. 
 
 
The prediction of thermoacoustic dynamics---even in simple systems---is challenging because of three reasons.
First, thermoacoustics is a multi--physics phenomenon. 
For a thermoacoustic oscillation to occur, three physical subsystems (flame, acoustics and hydrodynamics) constructively interact with each other~\citep[e.g.,][]{lieuwen_book_2012,magri_adjoint_2019}.
Second, thermoacoustics is a nonlinear phenomenon~\citep[e.g.,][]{sujith_pof_2020}. 
In general, the flame's heat release responds nonlinearly to acoustic perturbations \citep{dowling_kinematic_1999}; and 
the hydrodynamics are typically turbulent~\citep[e.g.,][]{huhn_stability_2019}.  
Third, thermoacoustics is sensitive to perturbations to the system.
In the linear regime, small changes to the system's parameters, such as the flame time delay, can cause arbitrarily large changes of the eigenvalue growth rates at exceptional points~\citep{mensah2018exceptional,orchini2020degenerate}.
In the nonlinear regime, small changes to the system's parameters can cause a variety of nonlinear bifurcations of the solution.
As a design parameter is varied in a small range, 
thermoacoustic oscillations may become chaotic, by either period doubling, or Ruelle--Takens–Newhouse scenarios~\citep{gotoda_dynamic_2011, gotoda_short_2012, kabiraj_nonlinear_2012, kashinath_nonlinear_2014, orchini_frequency_2015, huhn_stability_2019}, 
or by intermittency bifurcations scenarios \citep{nair_intermittency_2014,nair_reduced_2015}. 
The rich bifurcation behaviour has an impact on the effectiveness of control strategies, which may work for periodic oscillations with a dominant frequency, but may not work for multi--frequency oscillations as effectively. 
Additionally, unpredictable changes in the operating conditions and turbulence, which can be modelled as random phenomena~\citep{nair_reduced_2015,noiray2017linear}, increase the uncertainty on the prediction of the quantities of interest. 

 
Thermoacoustics can be modelled with a hierarchy of assumptions and computational costs.
Large--eddy simulations make assumptions only on the finer flow scales, which makes the final solution high--fidelity, but computationally expensive~\citep{poinsot_prediction_2017}. 
Euler and Helmholtz solvers compute the acoustics that evolve on a prescribed mean flow, which makes the solution medium--fidelity and computationally less expensive than turbulent simulations~\citep[e.g.,][]{nicoud_acoustic_2007}. 
This is commonly achieved with flame models, which capture the heat--release response to acoustic perturbations with transfer functions~\citep[e.g.,][]{noiray2008unified,silva_combining_2013} {and distributed time--delays~\citep{polifke2020modeling}.}
Other medium--fidelity and medium--cost methods are based on flame--front tracking~\citep[e.g.,][]{pitsch_large_2002} and simple chemistry models~\citep[e.g.,][]{magri2014global}, to name only a few. 
On the other hand, low--order models based on travelling waves and standing waves~\citep{dowling1995calculation} provide low--fidelity solutions, but with low--computational cost. 
These low--order models capture only the dominant physical mechanisms, which are the flame time delay, the flame strength (or index) and the damping. 
Low--order models, which are the subject of this study, are attractive to practitioners because they provide quick estimates on the quantity of interest.
%
%
Along with modelling, accurate experimental data is becoming more accessible and available~\citep{o2015transverse}. 
To monitor the thermoacoustic behaviour, in both real engines and academic rig (such as the Rijke tube), the pressure is experimentally measured by microphones 
\citep{lieuwen_combustion_2005, kabiraj_route_2012}. 
Microphones sample the pressure amplitude at typically high rates, which generates large datasets in real time. %
{Except when required for diagnostics and a--posteriori parameter identification~\citep[among many others,][]{polifke2001reconstruction,schuermans2003modeling,lieuwen_combustion_2005, noiray2008unified, noiray2017linear, polifke2020modeling}, the data is useful if it can be used in {\it real time}, i.e., on-the-fly, to correct (or update) our knowledge of the thermoacoustic states.   The sequential assimilation method that we develop bypasses the need to store data, which enables the real-time assimilation of data as well as on-the-fly parameter estimation. 
}\\ 

%
 
%
To summarise, in thermoacoustics, we have three ingredients to improve the design:
(i) a human being, who identifies the physical mechanisms that need to be modelled depending on the objectives and resources; 	
(ii) a mathematical model, which provides estimates of the physical states;  
and (iii) experimental data, which provides a quantitative measure of the system's observables. 
A model is \textit{good} if the human being identifies the physical mechanisms needed to formulate a mathematical model that provides the system's states compatibly with the experimental data.  
The overarching objective of this paper is to propose a method {\it to make qualitatively low--order models  quantitatively (more) accurate} every time that reference data becomes available. 
The ingredients for this are a physical low--order model, which provides the states; data, which provides the observables; and a statistical method, which finds the most likely model by assimilating the data in the model.  
In weather forecasting, this process is known as data assimilation~\citep{sasaki_fundamental_1955}. 
Data assimilation techniques have been applied to oceanographic studies~\citep{eckart_hydro_1960}, aerospace control~\citep{gelb_applied_1974}, robotics, geosciences, cognitive sciences~\citep{reich_probabilistic_2015}, to name only a few. 
Data assimilation is a principled method, which, in contrast to traditional machine learning, uses a physical model to provide a prediction on the solution (the \textit{forecast}), which is updated when observations become available to provide a corrected state (the \textit{analysis})~\citep{reich_probabilistic_2015}.  
The analysis is an estimator of the physical state (the \textit{true} state), which is more accurate than the forecast. 
%
%
\subsection{Data assimilation}
Data assimilation methods can be divided into two main approaches~\citep{lewis_assimilation_2006}: 
(i) variational  
and (ii) statistical assimilation methods. 
Variational data assimilation requires the minimisation of a cost functional---e.g., a Mahalanobis (semi)norm---in terms of a control variable to obtain a single optimal analysis state~\citep{bannister_review_2017}. 
The most common variational methods are 3D--VAR and 4D--VAR, which are widely used in weather centres such as the Met Office in the UK or the European  Centre for Medium--Range Weather Forecasts, and in academic research~\citep{bannister_review_2008}. 
In thermoacoustics, variational data assimilation was introduced by~\citet{traverso_data_2019}, who found the optimal thermoacoustic states given reference data from pressure observations. {Because variational methods need batches of data, they are not naturally suited to real--time inference. }
On the other hand, statistical data assimilation combines concepts of probability and estimation theory. 
The aim of statistical data assimilation is to compute the probability distribution function  of a numerical model to  statistically combine it with data from observations. 
Because the probability distribution function is high dimensional, the practitioner is often interested in capturing only the first and second statistical moments of it. 
In reduced--order modelling, this was achieved in flame tracking methods by~\citet{yu_combined_2019}, who implemented ensemble Kalman filters and smoothers to learn the flame parameters on the fly. 
In high--fidelity methods in reacting flows, data assimilation with ensemble Kalman filters have been applied in large--eddy simulation of premixed flames to predict local extinctions in a jet flame~\citep{labahn_data_2019}, and under--resolved turbulent simulation to predict autoignition events~\citep{magri_doan_2019}. 
The ensemble Kalman filter has also been successfully applied to non--reacting flow systems that show high nonlinearities such as the estimation of 
turbulent near--wall flows~\citep{colburn_state_2011}, 
uncertainties in Reynolds--averaged
Navier--Stokes (RANS) equations~\citep{xiao_quantifying_2016},
aerodynamic flows~\citep{da_ensemble_2018}. 
Statistical data assimilation based on Bayesian methods, {which enable real--time prediction in contrast to variational methods}, was introduced in thermoacoustics by~\citet{novoa_bayesian_2020}. 

\subsection{Model bias}
{
Data assimilation methods are commonly derived under the assumption that forecast errors are random with zero mean~\citep{evensen_data_2009}, or, in other words, the error is unbiased. 
However, in addition to state and parameter uncertainties, low--order models are  affected by model uncertainty, which manifests as an error bias. 
Modelling the model bias is an active research area. 
To produce an unbiased analysis, both forecast and observation biases need to be estimated~\citep{dee_data_1998}. 
 \citet{friedland_treatment_1969} developed the Separate Kalman Filter to estimate the bias, which is a two--stage sequential filtering process that addresses the estimation of a constant bias term, but its application is limited to linear processes. 
\citet{drecourt_bias_2006} extended the implementation of the Separate Kalman Filter and compared it to the Coloured Noise Kalman Filter, which augments the state vector with an auto--retrogressive model that describes the bias. They propose a feedback implementation of these methods, which allows a time--correlated representation of the bias, but the accuracy is limited by the prescribed model of the bias.  
More recently, \citet{dasilva_flow_2020} proposed a low--rank representation of the observation discretisation  bias, based as well on an auto--regressive model. They performed parameter estimation with an Ensemble Kalman Filter to calibrate the parameters of the auto--regressive model. This successfully modelled the discretisation bias, but did not tackle the model bias, which is a more general form of error. 
The estimation of a nonlinear dynamical state in the presence of a model bias remains an open problem. 
We propose a framework to obtain an unbiased analysis in thermoacoustic low--fidelity models by inferring the model bias.
This consists of the combination of data assimilation with a recurrent neural network, which infers the model error of the low--fidelity of the thermoacoustic system.

Recurrent neural networks are data--driven techniques that are designed to learn temporal correlations in time series~\citep{rumelhart_learning_1986}, with a variety of applications in timeseries forecasting.  
In fluid mechanics, recurrent neural networks have been employed to model unsteady flow around bluff bodies~\citep{hasegawa_machine_2020}, and as an optimisation tool for gliding control~\citep{novati_controlled_2019}. 
Among recurrent neural networks, echo state networks based on reservoir computing, which are  universal approximators~\citep{GRIGORYEVA2018495}, proved to be successful in learning nonlinear correlations in data~\citep{maass2002real,jaeger2004harnessing} and ergodic properties~\citep{huhn_gradient_2022}.   
Training an echo state network consists of a simple linear regression problem. Because no gradient descent is necessary, vanishing/exploding gradient problems do not occur in echo state networks.  
In chaotic flows, echo state networks have been used, for instance, to learn and optimise the time average of thermoacoustic dynamics~\citep{huhn_learning_2020, huhn_gradient_2022}, to predict turbulent dynamics with physical constraints~\citep{doan2021short}, and to predict the statistics of extreme events in turbulent flows \citep{racca_robust_2021}. 
In this paper, we propose to model the model bias with an echo state network, which is a tool that is more versatile and general than auto--regressive models~\citep[p.306,][]{aggarwal2018neural}. 
}

\subsection{Objectives and structure}
The objective of this paper is five--fold. 
First, we develop a sequential data assimilation for a low--order model to self--adapt and self--correct any time that reference data becomes available. The method, which is based on Bayesian inference, provides the \textit{maximum a posteriori estimate} model prediction, i.e., the most likely prediction. 
Second, we apply the methodology to infer the thermoacoustic states and heat release parameters on the fly without storing data. 
Third, we analyse the performance of the data assimilation algorithm on a twin experiment with synthetic data and interpret the results physically. 
Fourth, we propose practical rules for thermoacoustic data assimilation. 
{Fifth, we extend the data assimilation method to account for a biased thermoacoustic model. 
This method is tested by assimilating observations from a higher--fidelity model with non--ideal boundary conditions, a mean flow and the simulation of the flame front with a kinematic model~\citep{dowling_kinematic_1999}. The simulation of the flame dynamics is suitable for a time--domain approach, and it overcomes the limitations of flame response models.} 

 %
The paper is structured as follows. 
\S~\ref{sec:generalfra} provides a description of the nonlinear thermoacoustic model with the data assimilation technique and its implementation for thermoacoustics.
\S~\ref{sec:method} presents the method developed for state and parameter estimation. 
{\S~\ref{sec:bias_estimation} presents the method developed for combining data assimilation with an echo state network.  }
\S~\ref{sec:characterisation} presents the nonlinear characterisation of the thermoacoustic dynamics.  
\S~\ref{sec:results} shows the results for non--chaotic regimes, whereas 
\S~\ref{sec:CH_results} shows and discusses the results for chaotic solutions. 
{\S~\ref{sec:bias_results} shows and discusses the results for unbiased state and parameter estimation for a high--fidelity limit cycle. }
A final discussion and conclusions end the paper in~\S~\ref{sec:conclusions}.
 
\section{Thermoacoustic data assimilation}\label{sec:generalfra}
We consider a nonlinear thermoacoustic model, $\mathcal{T}$, as 
\begin{align}
\mathcal{T}\left(\boldsymbol{\psi}, \boldsymbol{\alpha}, \mathbf{y},  \mathbf{F}, \boldsymbol{\delta} \right)& = 0, \\
\mathcal{G}(\boldsymbol{\psi}) & = \mathbf{y}  + \boldsymbol{\epsilon}
\end{align}
where 
$\boldsymbol{\psi}$ is the state of the system; 
$\boldsymbol{\alpha}$ is the vector of the system's parameters;  
$ \mathbf{y}$ are the observables from reference data; 
$\mathbf{F}$ is a nonlinear operator, which, in general, is differential; 
$\boldsymbol{\delta}$ is a model error; 
$\mathcal{G}$ is a nonlinear map from the state space to the observable state; 
and $\boldsymbol{\epsilon}$ is the observation error. 
The thermoacoustic problem  on which we focus is: 
{\it Given some data (observables) $\mathbf{y}$, and a mathematical model $\mathcal{T}$, what are the most likely physical states $\boldsymbol{\psi}$ and parameters, $\boldsymbol{\alpha}$, of the system?}
To answer this, we use a Bayesian approach in the well--posed maximum a posteriori estimation framework.
Although the framework is versatile, in the next sections, we specify the low--order model, $\mathbf{F}$, the data, $\mathbf{y}$, and the data assimilation approach. 
%


\subsection{Qualitative nonlinear thermoacoustic model}\label{sec:Rijke}

\begin{figure}
    \centering
    \includegraphics[width=.7\textwidth]{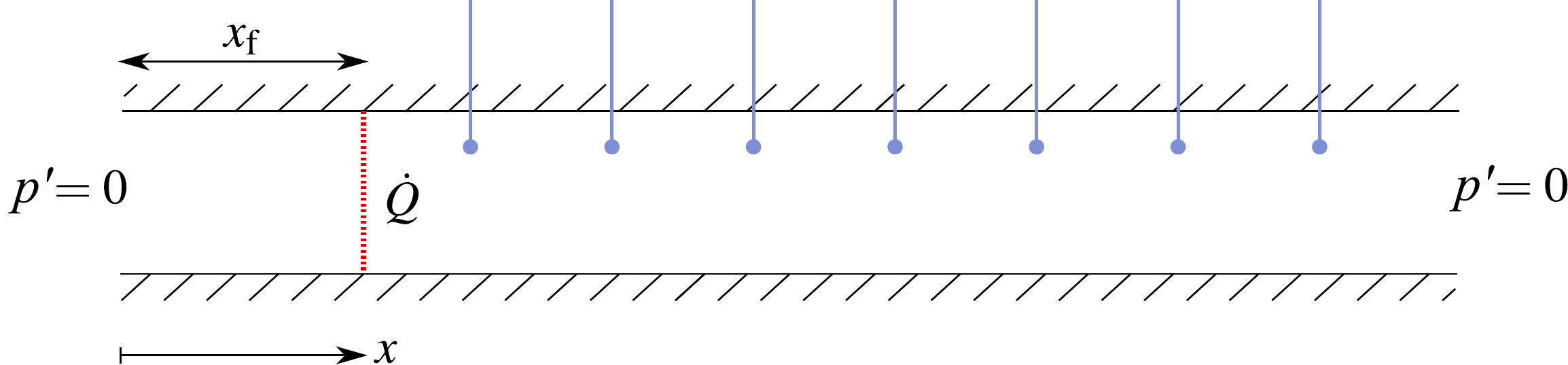}
    \caption[Schematic of a Rijke tube]{Schematic of an open--ended duct with a heat source (also known as the Rijke tube). The heat released  by the compact heat source is indicated by the vertical dotted line. The light blue vertical lines indicate microphones located equidistantly.}
    \label{fig:rijke_tube}
\end{figure}

The system consists  of an open--ended tube containing a heat source, such as a flame or an electrically heated gauze. 
Because the tube is sufficiently long with respect to the diameter, the cut--on frequency is such that only longitudinal acoustic waves propagate. 
This is known as the Rijke tube, which is a common laboratory--scale device that has been employed in a variety of fundamental studies~\citep{heckl1990,balasubramanian_thermoacoustic_2008, juniper2011, magri_jfm_2013}. This device is represented in Figure~\ref{fig:rijke_tube}.
The Rijke model used in this work is described by \citet{balasubramanian_thermoacoustic_2008} and \citet{juniper2011}. 
The flow is assumed to be a perfect gas; 
the mean flow is sufficiently slow such that its effects are neglected in the acoustic propagation; and 
 viscous and body forces are neglected. 
The acoustics are governed by the dimensionless linearised  momentum and energy conservation equations  

\begin{subequations}\label{eq:final_syst}
\begin{align}
    \label{eq:final_syst2}
    &\dfrac{\partial{u}'}{\partial{t}}+\dfrac{\partial{p}'}{\partial{x}}=0\\
    \label{eq:final_syst3}
    &\dfrac{\partial{p}'}{\partial{t}}+\dfrac{\partial{u}'}{\partial{x}}=\Dot{Q}\,\delta\,({x}-{x}_\mtxt{f}) - \zeta\,{p}'
\end{align}
\end{subequations}
where 
$u'$ is the acoustic velocity;  
$p'$ is the acoustic pressure; 
$\Dot{Q}$ is the heat release rate; 
$x_\mathrm{f}$ is the flame location; 
${\delta}$ is the Dirac delta distribution, which models the heat source as a point source (compact assumption); 
and $\zeta$ is the damping factor, which encapsulates the acoustic energy radiated from both ends of the duct, and the thermo--viscous losses in boundary layers. 
The non--dimensional heat release rate perturbation, ${\Dot{Q}}$, is modelled with a qualitative nonlinear time--delayed model~\citep{heckl1990}
\begin{equation}
\label{eq:heat_source}
    {\Dot{Q}}= \beta\,\left[\sqrt{\left|\tfrac{1}{3}+{u'}_\mtxt{f}({t}-{\tau})\right|}-\sqrt{\tfrac{1}{3}}\right]
\end{equation}
where $\beta$ is the strength of the source;  
$u'_\mtxt{f}$ is the acoustic velocity at the flame location; 
and ${\tau}$ is the time delay. 
The heat release rate is a key thermoacoustic parameter for the system's stability. 
The dimensionless variables in \eqref{eq:final_syst}-\eqref{eq:heat_source} and the dimensional variables (with~$\,\tilde{\;}\,$) are related as 
$x = {\tilde{x}}/{\tilde{L}_0}$, where ${\tilde{L}_0}$ is the length of the tube;   
$t = {\tilde{t}}{\tilde{c}_0}/\tilde{L}_0$, where $\tilde{c}_0$ is the mean speed of sound;  
$u' = {\tilde{u}'}/{\tilde{c}_0}$; 
$\rho' = {\tilde{\rho}'}/{\tilde{\rho}_0}$, where $\tilde{\rho}_0$ is the mean density;  
$p' = {\tilde{p}'}/({\tilde{\rho}_0\,\tilde{c}_0^2})$;  
$\Dot{Q} = {\tilde{\Dot{Q}}'\,(\gamma-1)}/({\tilde{\rho}_0\,\tilde{c}_0^3})$, where $\gamma$ is the heat capacity ratio;  
and $\delta({x}-{x}_\mtxt{f}) = {\tilde{\delta}(\tilde{x}-\tilde{x}_\mtxt{f})}{\tilde{L}_0}$. 
The open--ended boundary conditions are ideal, which means that the acoustic pressure is zero, i.e., $p'=0$ at $x=\{0,1\}$.  
By separation of variables, the acoustic velocity and pressure are decomposed into Galerkin modes as~\citep{zinn_application_1971}
\begin{equation}
    \label{eq:decomposition}
    u'(x,t)=\sum^{N_m}_{j=1}\,\eta_j(t)\,\cos{(j\pi x)}\;,\qquad
    p'(x,t)=-\sum^{N_m}_{j=1}\,\dfrac{\Dot{\eta}_j(t)}{j\,\pi}\,\sin{(j\pi x)}
\end{equation}
where $\cos(j\pi x)$ and $\sin(j\pi x)$ are the eigenfunctions of the acoustic velocity and pressure, respectively, when $\zeta=0$ and $\dot{Q}=0$; 
and $N_m$ is the number of acoustic modes kept in the decomposition. 
Substituting \eqref{eq:decomposition} into \eqref{eq:final_syst}, multiplying \eqref{eq:final_syst3} by $\sin{(k\pi x)}$, and integrating over $x=[0,1]$, yield the governing ordinary differential equations, which physically represent a set of nonlinearly coupled oscillators 
\begin{subequations}
\label{eq:discrete}
\begin{align}
\label{eq:discrete_1}
    \dfrac{d \eta_j}{d t}-j\pi\left(\dfrac{ \Dot{\eta}_j}{j\,\pi}\right)&=0 \\[0.3em]\label{eq:discrete_2}
    \dfrac{d}{d t}\left(\dfrac{\Dot{\eta}_j}{j\,\pi}\right) +j\pi\,\eta_j + \zeta_j\,\dfrac{\Dot{\eta}_j}{j\,\pi}+2\,\Dot{Q}\,\sin{(j\pi x_\mtxt{f})}&=0
\end{align}
\end{subequations}
where the damping term is defined by modal components $\zeta_j=C_1\,j^2+C_2\sqrt{j}$, which is physically motivated in~\citet{LANDAU_1987}. 
The damping coefficients, $C_1$ and $C_2$, are assumed to be constant. 
For reasons that will be explained in~\S~\ref{sec:DA}, we introduce an advection equation to mathematically eliminate the time--delayed velocity term~\citep{huhn_stability_2019} 
\begin{equation}
    \label{eq:advection}
    \dfrac{\partial v}{\partial t} + \dfrac{1}{\tau}\dfrac{\partial v}{\partial X} = 0 \quad,\qquad0\leq X\leq1
\end{equation}
where $v$ is a dummy variable that travels with non--dimensional velocity $\tau^{-1}$ in a dummy spatial domain $X$ such that 
\begin{equation}    \label{eq:advection_BCa}
u'_\mtxt{f}\left(t-\tau\right) = v\left(X = 1,\,t\right), \qquad       
     u'_\mtxt{f}\left(t\right)=v\left(X = 0,\,t\right). 
\end{equation}
Equation \eqref{eq:advection_BCa} is discretised with a Chebyshev method~\citep{trefethen_spectral_2000} with  $N_c + 1$ points in the interval $0\leq X\leq1$. 

In a state--space notation, the thermoacoustic problem is governed by 
\begin{align}
\frac{d\vect{\psi}}{dt} &= \mathbf{F}\left(\vect{\alpha}; \vect{\psi}\right),  \qquad 
\vect{\psi}(t=0) = \vect{\psi}_0\label{eq:ivp_compact}, \nonumber \\ 
\mathbf{y}
&= \mathbf{M}(x)\boldsymbol{\psi}, 
\end{align}
where the state vector $\vect{\psi}\equiv(\boldsymbol{\eta};\boldsymbol{\Dot{\eta}};\vect{v})\in\mathbb{R}^{2N_m+N_c}$ is the column--concatenation of the acoustic amplitudes, $\boldsymbol{\eta} \equiv (\eta_1,\eta_2,...,\eta_{N_m})\in\mathbb{R}^{N_m}$ 
and $\boldsymbol{\Dot{\eta}} \equiv ({\Dot{\eta}_1}/{\pi},{\Dot{\eta}_2}/{(2\,\pi)},...,{\Dot{\eta}_{N_m}}/{(N_m\,\pi)})\in\mathbb{R}^{N_m}$, 
and the dummy velocity variables $\vect{v}\equiv(v_1, v_2, ..., v_{N_c})\in\mathbb{R}^{N_c}$, {which arise from the discretisation of \eqref{eq:advection}}; 
the thermoacoustic parameters are contained in the vector $\vect{\alpha}=(\beta, \tau, \zeta) \in\mathbb{R}^{N_P}$; 
 $\mathbf{F}$ represents the nonlinear operator that consists of \eqref{eq:discrete_1},\eqref{eq:discrete_2} and \eqref{eq:advection}, $\mathbf{F}: \mathbb{R}^{2N_m+N_c+N_P}\rightarrow \mathbb{R}^{2N_m+N_c}$; and 
 $\mathbf{M}(x)$ is the measurement operator, which maps the state to the observable space at $x$. The expression of the measurement operator  depends on the nature of the observables being assimilated, as explained in \S~\ref{sec:method}.  
To work with a reduced--order model that {\it qualitatively} captures the essential dynamics, we use $N_m=10$ acoustic modes. 
For the advection equation, $N_c = 10$ ensures numerical convergence~\citep{huhn_stability_2019}. 
The number of degrees of freedom of the reduced--order model is $N= 2\,N_m + N_c=30$. 
The initial value problem \eqref{eq:ivp_compact} is solved with an automatic--stepsize--control method that combines fourth and fifth order Runge--Kutta methods~\citep{Shampine1997}.

\subsection{Qualitative and quantitative accuracy}
{
We say that a low--order model is {\it qualitatively} correct when it captures the key physical parameters/mechanisms (e.g., the time delay). 
Although a low--order model may be physically motivated, 
it is subject to three sources of errors: 
(i) uncertainty in the state, (ii) uncertainty in the parameters, and (iii) bias in the model, i.e., the low--order equation does not contain all the terms necessary to model the phenomenon (the model bias is equivalently referred to as the model error). Data assimilation methods combine the forecast of a low--order model with observations from either real experiments, or high--fidelity simulations, which reduces the bias in the state (\S~\ref{sec:state_estimation}) and/or in the parameters of the model (\S~\ref{sec:Parameter_estimation}). However, traditional data assimilation methods do not tackle the model bias because they assume that the forecast model is unbiased. In \S~\ref{sec:bias_estimation}, we propose an echo state network as a method to estimate the model bias, thereby closing the low--order model equations in the data assimilation. In summary, the aim of data assimilation is to make a \textit{qualitative} accurate model more \textit{quantitatively} correct. 
}

\subsection{Data assimilation}\label{sec:DA}

Data assimilation optimally combines 
the prediction from an imperfect model with data from observations to improve the knowledge of the system's state. 
The updated solution (\textit{analysis}) optimally combines the information from the observations, $\mathbf{y}$, and the model solution (\textit{forecast}) with their uncertainties.
In order to 
(i) update the system's knowledge any time that data becomes available, and 
(ii) not store the data during the entire operation, 
we assimilate sequentially assuming that the process is a Markovian process. 
The concept of Bayesian update is key to this process, as explained in~\S~\ref{sec:ffffe4}. 

\subsubsection{Bayesian update} \label{sec:ffffe4}

In a Bayesian framework, we quantify our confidence in a model by a probability measure.
Hence, we update our confidence in the model predictions every time we have reference data from observations.
The rigorous framework to achieve this is probability theory, as explained in Cox's theorem~\citep{jaynes_probability_2003}. 

To set a probabilistic framework at time $t=t_k$, the state, $\boldsymbol{\psi}_k$, and reference observation, $\mathbf{y}_k$ are assumed to be realisations of their corresponding random variables acting on the sample spaces $\Omega_{\boldsymbol{\psi}}=\mathbb{R}^{2N_m+N_c}$ and $\Omega_{\mathbf{y}}=\mathbb{R}^{N_{\mathbf{y}}}$.  
Because we transformed the time--delayed problem into an initial value problem,  the solution of~\eqref{eq:ivp_compact} at the present depends on the solution at the previous time step only. 
In other words, we transformed a non--Markovian system into a Markovian system, which simplifies the design of the Bayesian update. 
We quantify our confidence in a quantity through a probability, $\mathcal{P}$ 
\begin{equation} \label{eq:ff33}
\boldsymbol{\psi}_k \sim \mathcal{P}(\boldsymbol{\psi}_k | \boldsymbol{\psi}_{k-1}, \boldsymbol{\alpha}, \mathbf{F})\qquad 
\mathbf{y}_k \sim \mathcal{P}(\mathbf{y}_k | \boldsymbol{\psi_{k}}, \boldsymbol{\alpha}, \mathbf{F}), 
\end{equation}
where $\lvert$ denotes that the quantity on the left is conditioned on the knowledge of the quantities on the right.
The leftmost probability answers the question:
``Given a model $\mathbf{F}$, a set of parameters $\boldsymbol{\alpha}$, and the state $\boldsymbol{\psi}_{k-1}$, 
what is the probability that the state takes the value $\boldsymbol{\psi}_k$?''. 
The rightmost probability answers the question:
``if we forecast the state $\boldsymbol{\psi_k}$ from the model, what is the probability that we observe $\mathbf{y}_k$?''. 
We assume that the observations are statistically independent and uncorrelated with respect to the forecast. 
To update our knowledge of the system, the prior knowledge from the reduced--order model and the reference observations are combined through Bayes' rule 
\begin{align}\label{eq:bayup}
\mathcal{P}(\boldsymbol{\psi}_k | \mathbf{y}_k, \boldsymbol{\alpha}, \mathbf{F}) = \frac{\mathcal{P}(\mathbf{y}_k | \boldsymbol{\psi}_k, \boldsymbol{\alpha}, \mathbf{F}) \mathcal{P}(\boldsymbol{\psi}_k, \boldsymbol{\alpha}, \mathbf{F})}{\mathcal{P}(\mathbf{y}_k, \boldsymbol{\alpha}, \mathbf{F})}. 
\end{align}
First,  
$\mathcal{P}(\boldsymbol{\psi}_k, \boldsymbol{\alpha}, \mathbf{F})$ is the prior, which measures
the knowledge of our system prior to observing $\mathbf{y}_k$. 
The prior evolves through the Chapman--Kolmogorov equation~\citep{jazwinski2007stochastic}, which involves multi--dimensional integrals.
To numerically solve the Chapman--Kolmogorov equation, we use an ensemble method 
by integrating the model equations (\S~\ref{sec:gfojgf38f39}), which provide a {\it forecast} on the state. 
Second, 
$\mathcal{P}(\mathbf{y}_k | \boldsymbol{\psi}_k, \boldsymbol{\alpha}, \mathbf{F})$ is the likelihood~\eqref{eq:ff33}, 
which measures the confidence we have in our model prediction.
{The likelihood is prescribed (see \S~\ref{sec:gfojgf38f39})}. 
Third, 
$\mathcal{P}(\mathbf{y}_k, \boldsymbol{\alpha}, \mathbf{F})$ is the evidence, which is the probability that the observable takes on the value $\mathbf{y}_k$. 
This can be prescribed from the knowledge of the experimental uncertainties. 
Finally, 
$\mathcal{P}(\boldsymbol{\psi}_k | \mathbf{y}_k, \boldsymbol{\alpha}, \mathbf{F}) $ is the posterior, 
which measures the knowledge we have on the state, $\boldsymbol{\psi}_k$, after we have observed $\mathbf{y}_k$. 
Here, we will select the most probable value of $\boldsymbol{\psi}_k$ in the posterior (i.e., the mode) as the best estimator of the state 
({\it maximum a posteriori}  approach, which is a well--posed approach in inverse problems). 
The best estimator is called {\it analysis} in weather forecasting~\citep{tarantola2005inverse}.  
Equation~\eqref{eq:bayup} provides the Bayesian update, which is key to this work and sequential data assimilation. 
%

%
%
\subsubsection{Stochastic ensemble filtering for sequential assimilation}\label{sec:gfojgf38f39}

For brevity, we will omit the subscript $k$, unless it becomes necessary for clarity. 
We focus on a {\it qualitative} reduced--order model  in which 
(i) the bias on the solution is negligible,
 (ii) the uncertainty on the state is represented by a covariance, 
 (iii) the probability density function of the state is assumed to be symmetrical around the mean, and
 (iv) the dynamics at regime do not present frequent extreme events, i.e., the tails of the probability density function  are not heavy. {In section~\S\ref{sec:bias_estimation} we relax assumption (i) by introducing a methodology to estimate the bias of the solution, i.e., the model error.}

The probability distribution to employ is the distribution that maximises the information entropy~\citep{jaynes_information_1957}, which, in this scenario, is the Gaussian distribution. 
Therefore, the system's forecast and the observations are assumed to follow Gaussian distributions, i.e., 
$\vect{\psi}^\mtxt{f} \sim \mathcal{N}\left(\vect{\psi},\vect{C}^\mtxt{f}_{\psi\psi}\right)$ and $\vect{y} \sim \mathcal{N}\left(\vect{M}\vect{\psi},\vect{C}_{\epsilon\epsilon}\right)$, respectively, 
where $\mathcal{N}$ denotes the normal distribution with the first argument being the mean, and the second argument being the covariance matrix. 
The forecast and observation covariance matrices are $\vect{C}^\mtxt{f}_{\psi\psi}$ and $\vect{C}_{\epsilon\epsilon}$, respectively.  
If the dynamics were linear, the Bayesian update~\eqref{eq:bayup} would be exactly solved by the Kalman filter equations~\citep{kalman1960new} 
\begin{subequations} 
\label{eq:KF_analysis}
\begin{align}
    \label{eq:KF_psia}
     \vect{\psi}^\mtxt{a} &= \vect{\psi}^\mtxt{f} + \left(\vect{M}\,\vect{C}^\mtxt{f}_{\psi\psi}\right)^\mtxt{T}
                    \left[\vect{C}_{\epsilon\epsilon} + \vect{M}\,\vect{C}^\mtxt{f}_{\psi\psi}\,\vect{M}^\mtxt{T}\right]^{-1}\left(\vect{y}-\vect{M}\,\vect{\psi}^\mtxt{f}\right) \\[1em]
    \label{eq:KF_C}
    \vect{C}^\mtxt{a}_{\psi\psi} &= \vect{C}^\mtxt{f}_{\psi\psi} -
                    \left(\vect{M}\,\vect{C}^\mtxt{f}_{\psi\psi}\right)^\mtxt{T}
                    \left[\vect{C}_{\epsilon\epsilon} + \vect{M}\,\vect{C}^\mtxt{f}_{\psi\psi}\,\vect{M}^\mtxt{T}\right]^{-1}
                    \left(\vect{M}\,\vect{C}^\mtxt{f}_{\psi\psi}\right)
\end{align}
\end{subequations}
where the superscripts `a' and `f' denote analysis and forecast, respectively.  
Equation~\eqref{eq:KF_psia} corrects the model prediction by weighting the statistical distance between the observations (data) and the forecast, according to the prediction and observation covariances~\citep{evensen_ensemble_2003}. 
The observation error covariance has to be prescribed based on the knowledge of the experimental methodology used. 

\begin{figure}
    \centering
    \includegraphics[width=.6\textwidth]{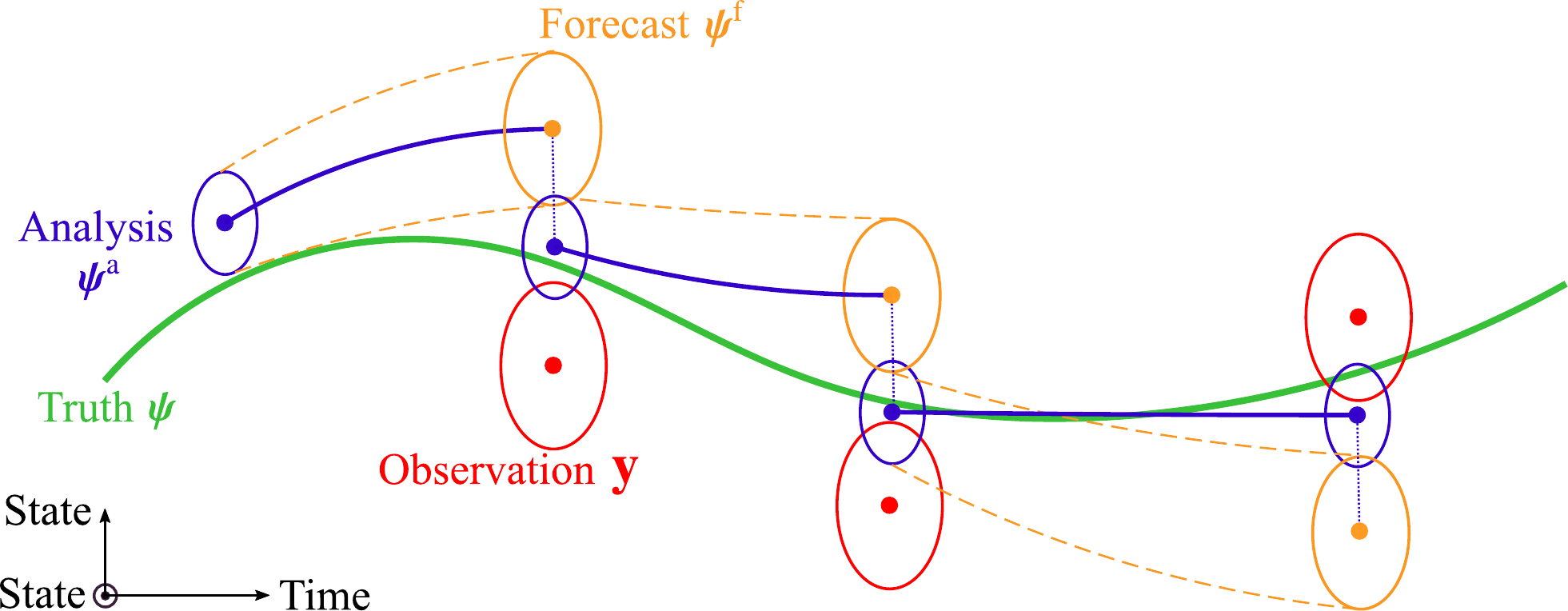}
    \caption{Conceptual schematic of a sequential filtering process. Truth (green); observations and their uncertainties (red); forecast state and uncertainties (orange); and analysis state and uncertainties (blue). The circles represent pictorially the spread of the probability density functions: the larger the circles, the larger the uncertainty.}
    \label{fig:filtering_process}
\end{figure}

In an ensemble method, the distribution is represented by the sample statistics
\begin{align}\label{eq:frhfwe}
\overbar{\boldsymbol{\psi}} \approx \frac{1}{m}\sum_{i=1}^m \boldsymbol{\psi}^i, \qquad \mathbf{C}_{\boldsymbol{\psi}\boldsymbol{\psi}} \approx \frac{1}{m-1}\boldsymbol{\Psi}\boldsymbol{\Psi}^T
\end{align}
where the $i$--th column of the matrix $\boldsymbol{\Psi}$ is the deviation from the mean of the $i$--th realisation, $\boldsymbol{\psi}^i - \overbar{\boldsymbol{\psi}}$, and 
$m$ is the number of ensemble members. 
 Because~\eqref{eq:frhfwe} is a Monte Carlo Markov Chain integration, the sampling error scales as~$\mathcal{O}(N^{-1/2})$. 
The key idea of ensemble filters is to group forecast states from a numerical model (the ensemble) to obtain, on filtering, the analysis state. 
Ensemble methods describe the state's uncertainty by the spread in the ensemble at a given time to avoid the explicit formulation of the covariance matrices~\citep{livings_unbiased_2008}. 
The algorithmic procedure is as follows.
First, the initial condition is integrated forward in time to provide the forecast state, $\vect{\psi}^\mtxt{f}$. 
Second, experimental observations, $\vect{y}$, are statistically assimilated into the forecast to obtain the analysis state, $\vect{\psi}^\mtxt{a}$, which, in turn, becomes the initial condition for the next time step. 
The forecast accumulates errors over the integration period, which is reduced in the assimilation stage through observations with their experimental uncertainties. 
If the model is qualitatively correct and unbiased, after a sufficient number of assimilations, the ensemble concentrates around the true value. 
This sequential filtering process on one ensemble member is shown in Figure~\ref{fig:filtering_process}.  
The process is repeated in parallel for the other ensemble members. 

%

\subsubsection{Ensemble Square--Root Kalman Filter} \label{sec:ensrkf}
In the ensemble Kalman filter~\eqref{eq:KF_analysis}, each ensemble member is updated with the assimilation  of independently perturbed observation data.
However, this method provides a sub--optimal solution that, in some cases, does not preserve the ensemble mean and is affected by sampling errors of the observations~\citep{evensen_ensemble_2003}. 
Moreover, the  ensemble Kalman filter may require a fairly large ensemble to compensate the sampling errors of the observations~\citep{sakov_implications_2008}. 
The ensemble square--root Kalman filter (EnSRKF), which is an ensemble--transform Kalman filter, overcomes these issues~\citep{livings_unbiased_2008}. 
The key idea of the EnSRKF is to update the  ensemble mean and deviations instead of each ensemble member. 
The EnSRKF for $m$ ensemble members and a state vector of size $N$ reads 
\begin{subequations}\label{th:EnSRKF}
\begin{align}
    \vect{A}^\mtxt{a} \;&=\;  \overbar{\vect{A}}^\mtxt{a} + \vect{\Psi}^\mtxt{a}\\
    \overbar{\vect{A}}^\mtxt{a} &= \overbar{\vect{A}}^\mtxt{f} + \vect{\Psi}^\mtxt{f}\,\left( {\vect{M}\,\vect{\Psi}^\mtxt{f}}\right)^\mtxt{T}
                    \left[(m-1)\,\vect{C}_{\epsilon\epsilon} +  {\vect{M}\,\vect{\Psi}^\mtxt{f}}\,\left( {\vect{M}\,\vect{\Psi}^\mtxt{f}}\right)^\mtxt{T}\right]^{-1}\left(\vect{Y}-\vect{M}\,\overbar{\vect{A}}^\mtxt{f}\right) \\
    \vect{\Psi}^\mtxt{a} \;&=\; \vect{\Psi}^\mtxt{f}\,\vect{V}\left(\mathbb{I}-\,\vect{\Sigma}\right)^{\rfrac{1}{2}}\vect{V}^\mtxt{T}\\
    \vect{V}\vect{\Sigma}\vect{V}^\mtxt{T} \;&=\; \left(\vect{M}\,\vect{\Psi}^\mtxt{f}\right)^\mtxt{T}\,\left[(m-1)\,\vect{C}_{\epsilon\epsilon} +  {\vect{M}\,\vect{\Psi}^\mtxt{f}}\,\left( {\vect{M}\,\vect{\Psi}^\mtxt{f}}\right)^\mtxt{T}\right]^{-1}\left(\vect{M}\,\vect{\Psi}^\mtxt{f}\right)
\end{align}
\end{subequations}
where $\vect{A} = (\vect{\psi}_1,\, \vect{\psi}_2,\, \dots,\, \vect{\psi}_m)\in \mathbb{R}^{N \times m }$ is the matrix that contains the ensemble members as columns; 
$\overbar{\vect{A}}= (\overbar{\boldsymbol{\psi}},\, \dots,\, \overbar{\boldsymbol{\psi}})\in\mathbb{R}^{N \times m}$ contains the mean analysis states in each column; 
$\vect{Y}=(\vect{y},\dots,\vect{y})\in \mathbb{R}^{q \times m}$ is the matrix containing the {$q$} observations repeated $m$ times;
the identity matrix is represented by $\mathbb{I}$; and  $\vect{V}$ and $\Sigma$ are the orthogonal matrices of eigenvectors and a diagonal matrix of eigenvalues, respectively,  from singular value decomposition. 
The largest matrices required in the EnSRKF algorithm  have dimension $N \times m$ and $m \times m$, therefore, the storage requirements are significantly smaller than those of non--ensemble based filters. 
 In addition, this filter is non--intrusive and suitable for parallel computation. 
A derivation of the EnSRKF can be found in Appendix~\ref{app:derivation}.

\subsection{Discussion}
An ensemble method enables us to
(i) work with high--dimensional systems because we do not need to propagate the covariance matrix, which has~$\mathcal{O}\,(N^2)$ components; 
(ii) work with nonlinear systems, such as the thermoacoustic system under investigation; 
(iii) work with time--dependent problems; 
(iv) not store the data because we sequentially assimilate (real--time, i.e., on--the--fly, assimilation); 
and 
(v) avoid implementing tangent or adjoint solvers, which are required, for example, in variational data assimilation methods~\citep{traverso_data_2019}. 
On the one hand, if the system were linear, a Gaussian prior would remain Gaussian under time integration. 
This makes the ensemble filter the exact Bayesian update in the limit of an infinite number of samples. 
On the other hand, if the system were nonlinear (e.g., in the present study), a Gaussian prior would not necessarily remain Gaussian under time integration. 
This makes the ensemble filter an approximate Bayesian update. 
The update of the first and second statistical moments, however, remains exact. 
In other words, we cannot capture the skewness, kurtosis, and other higher moments. 
(Particle filter methods overcome this limitation, but they may be computationally expensive~\citep{pham_stochastic_2001}.)

\section{State and parameter estimation}\label{sec:method}
This work considers both 
state estimation, in which the state is the uncertain quantity (\S~\ref{sec:state_estimation}); and combined state and parameter estimation, in which both the state and model parameters are uncertain (\S~\ref{sec:Parameter_estimation}).

\subsection{State estimation}\label{sec:state_estimation}
State estimation is the process of using a series of noisy measurements into an estimation of the state of the dynamical system, $\boldsymbol{\psi}$. 
This paper considers two different scenarios in assimilating acoustic data in thermoacoustics: 
(i) assimilation of the acoustic modes; 
and 
(ii) assimilation of pressure measurements from $N_\mtxt{mic}$ microphones, which are located equidistantly from the flame location up to the end of the Rijke tube (Figure~\ref{fig:rijke_tube}). 
The assimilation of acoustic modes assumes that observation data is available for the pressure and velocity acoustic modes, \{$\vect{\eta},\dot{\vect{\eta}}$\}. 
Hence, the state equations are
\begin{align}\label{eq:ivp_compact_SE_modes}
\frac{d\vect{\psi}}{dt}
&= 
\mathbf{F}\left(\vect{\alpha}; \vect{\psi}\right)
,  \qquad 
\vect{\psi}(t=0) = \vect{\psi}_0  =\begin{bmatrix}
\vect{\eta}_0  \\
\dot{\vect{\eta}}_0 \\
\vect{v}_0
\end{bmatrix} \nonumber \\
\mathbf{y}
&= \mathbf{M}(x)\boldsymbol{\psi}=\begin{bmatrix}
\vect{\eta}  \\
\dot{\vect{\eta}}
\end{bmatrix} 
\end{align}


%
Alternatively, in scenario (ii), from~\eqref{eq:decomposition}, the reference pressure measurements are computed as 
\begin{equation}
\label{eq:microphones}
\vect{p'_\mathrm{mic}} = \begin{pmatrix}
p'_1(t)           \\
p'_2(t)            \\
\vdots    \\   
p'_{N_\mtxt{mic}} (t)      \\
\end{pmatrix}
            = -
\begin{pmatrix}
\sin{\left(\pi x_1\right)} &\sin{\left(2\pi x_1\right)} &\dots &\sin{\left(N_m\pi x_1\right)}     \\
\sin{\left(\pi x_2\right)} &\sin{\left(2\pi x_2\right)} &\dots &\sin{\left(N_m\pi x_2\right)}     \\
\vdots &\vdots & &\vdots   \\
\sin{\left(\pi x_{N_\mtxt{mic}}\right) } &\sin{\left(2\pi x_{N_\mtxt{mic}} \right)} &\dots &\sin{\left(N_m\pi x_{N_\mtxt{mic}}\right) }      \\
            \end{pmatrix}
\begin{pmatrix}
\frac{\Dot{\eta}_1(t)}{\pi}     \\
\frac{\Dot{\eta}_2(t)}{2\pi}    \\
\vdots  \\
\frac{\Dot{\eta}_{N_m}(t)}{N_m\pi}     \\
            \end{pmatrix}
\end{equation}
The statistical errors of the microphones are assumed to be independent and Gaussian. 
%
In the twin experiment, the pressure observations are created from the true state, with a standard deviation $\sigma_\mtxt{mic}$ that mimics the measurement error. 
Pressure data cannot be assimilated directly with the EnSRKF because the state vector contains the acoustic modes, i.e., it does not contain the acoustic pressure. 
To circumvent this, we augment the state vector with the acoustic pressure at the microphones' locations according to \eqref{eq:microphones}.
Therefore, the new state vector includes the Galerkin acoustic modes, the {dummy velocity variables} and the pressure at the different microphone locations, i.e., $\vect{\psi}'\equiv(\boldsymbol{\eta};\boldsymbol{\Dot{\eta}};\vect{v};\vect{p'_\mathrm{mic}})$, with dimension $N' = 2\,N_m + N_c + N_\mathrm{mic}$. The augmented state equations are
\begin{align}\label{eq:ivp_compact_SE_mics}
\frac{d\vect{\psi'}}{dt}
&= 
\mathbf{F}\left(\vect{\alpha}; \vect{\psi}\right)
,  \qquad 
\vect{\psi'}(t=0) = \vect{\psi'}_0  =\begin{bmatrix}
\vect{\eta}_0  \\
\dot{\vect{\eta}}_0 \\
\vect{v}_0\\
\vect{p'_{\mathrm{mic}}}_0
\end{bmatrix} \nonumber \\
\mathbf{y}
&= \mathbf{M}(x)\boldsymbol{\psi'}=\vect{p'_{\mathrm{mic}}}(x)
\end{align}
With this, the modes will be updated indirectly during the assimilation step using the microphone data and their experimental error. 
%
%
%
%

\subsection{Combined State and Parameter estimation}\label{sec:Parameter_estimation}

Combined state and parameter estimation is the process of using a series of noisy measurements into an estimation of the state of the dynamical system, $\boldsymbol{\psi}$, and the parameters, $\boldsymbol{\alpha}$. 
The parameters are regarded as variables of the dynamical system so that they are updated in every analysis step. 
This is achieved by combining the governing equations of the thermoacoustic model 
with the equations that describe the evolution of parameters, which  are constant in time, but can change when observations are assimilated. 
The equations for the augmented state of combined state and parameter estimation are 
\begin{align}
\frac{d}{dt} 
\begin{bmatrix}
\vect{\psi}\\
\vect{\alpha}
\end{bmatrix}
&= 
\begin{bmatrix}
\mathbf{F}\left(\vect{\alpha}; \vect{\psi}\right)\\
0
\end{bmatrix}
,  \qquad 
\begin{matrix}
\vect{\psi}(t=0) = \vect{\psi}_0\\
\vect{\alpha}(t=0) = \vect{\alpha}_0
\end{matrix}\label{eq:ivp_compact_PE}, \nonumber \\ 
\mathbf{y}
&= \mathbf{M}(x)\boldsymbol{\psi}, 
\end{align}
With a slight abuse of notation, the state vector $\vect{\psi}$ in \eqref{eq:ivp_compact_PE}  is equal to $\vect{\psi}\equiv(\boldsymbol{\eta};\boldsymbol{\Dot{\eta}};\vect{v})$ in \eqref{eq:ivp_compact_SE_modes} for the assimilation of acoustic modes, and equal to $\vect{\psi'}\equiv(\boldsymbol{\eta};\boldsymbol{\Dot{\eta}};\vect{v};\vect{p'_\mtxt{mic}})$ in \eqref{eq:ivp_compact_SE_mics} for the assimilation of pressure measurements. 
The data assimilation algorithm is applied to the augmented system for both the forecast state and the parameters to be updated at every analysis step. 
The parameters need to be initialised for each ensemble member from a uniform distribution with a width of 25\% of the mean parameter value. 
In other words, we assume that the parameters are uncertain by $\pm 25\%$. \\  


\subsection{Performance metrics}
The performance of the state estimation and combined state and parameter estimation are evaluated with three metrics: 
(i) the trace of the forecast covariance, $\vect{C}_{\psi\psi}^\mtxt{f}$, which globally measures the spread of the ensemble;   
(ii) the relative difference between the true pressure oscillations at the flame location and the filtered solution, which measures the instantaneous error; 
and
(iii) for the combined state and parameter assimilation, the convergence of the filtered parameters normalised to their true values, as well as the root--mean square error with respect to the true solution.

{
\section{Data assimilation with bias estimation}\label{sec:bias_estimation}
In this section, we analyse the case of state, parameter and model bias estimation.
%
%
Sources of model bias in the model of~\S~\ref{sec:Rijke} include 
(i) idealised boundary conditions, 
(ii)  simple heat--release law with no simulation of the flame, and 
(iii) zero mean flow effects.  
We infer the model bias  to correct the biased forecast state prior to the analysis step. By performing state and parameter estimation on the unbiased forecast, we increase the quantitative accuracy of the model prediction.
First, we define the time--dependent model bias $\vect{U}(t)$ as the difference between the true pressure state (from the higher-fidelity model) at the microphone locations $\vect{{p}'^{~t}_{\mathrm{mic}}}$, and the expected biased pressure $\left\langle\vect{{p}'_{\mathrm{mic}}}\right\rangle$, i.e., the mean of the ensemble of pressures
\begin{equation}\label{eq:bias}
    \vect{U}(t) = \vect{{p}'^{~t}_{\mathrm{mic}}}(t) - \left\langle\vect{{p}'_{\mathrm{mic}}}(t)\right\rangle. 
\end{equation}
%
We propose an echo state network to predict the evolution of the model bias.

\subsection{Echo State Networks} 
An Echo State Network (ESN) is a type of recurrent neural network based on reservoir computing~ \citep{lukovsevivcius_practical_2012}. 
ESNs learn temporal correlations in data by nonlinearly expanding the data into a high--dimensional reservoir, which acts as the memory of the system, and is a framework that is more versatile than auto--regressive models~\citep{aggarwal2018neural}.

Figure~\ref{fig:ESN} shows a pictorial representation of an echo state network. 
The reservoir is defined by a high--dimensional vector $\textbf{r}(t_i)\in\mathbb{R}^{N_{r}}$, and a state matrix $\mathbf{W}\in\mathbb{R}^{N_{r}\times N_{r}}$, where $N_{r}$ is the number of neurons in the reservoir. We use $N_{r} = 100$ neurons, which are sufficient to represent the dynamics of the thermoacoutic system~\citep{huhn_gradient_2022}. 
The inputs and outputs from the reservoir are vectors of dimension $\mathbb{R}^{N_\mtxt{mic}}$ because we define the bias as the pressure error at each microphone~\eqref{eq:bias}. 
Their input and output matrices are $\mathbf{W}_{\mathrm{in}}\in\mathbb{R}^{N_{r}\times (N_\mtxt{mic}+1)}$ and $\mathbf{W}_{\mathrm{out}}\in\mathbb{R}^{ N_\mtxt{mic}\times(N_{r}+1)}$, respectively.
\begin{figure}
    \centering
    \includegraphics[width=.4\linewidth]{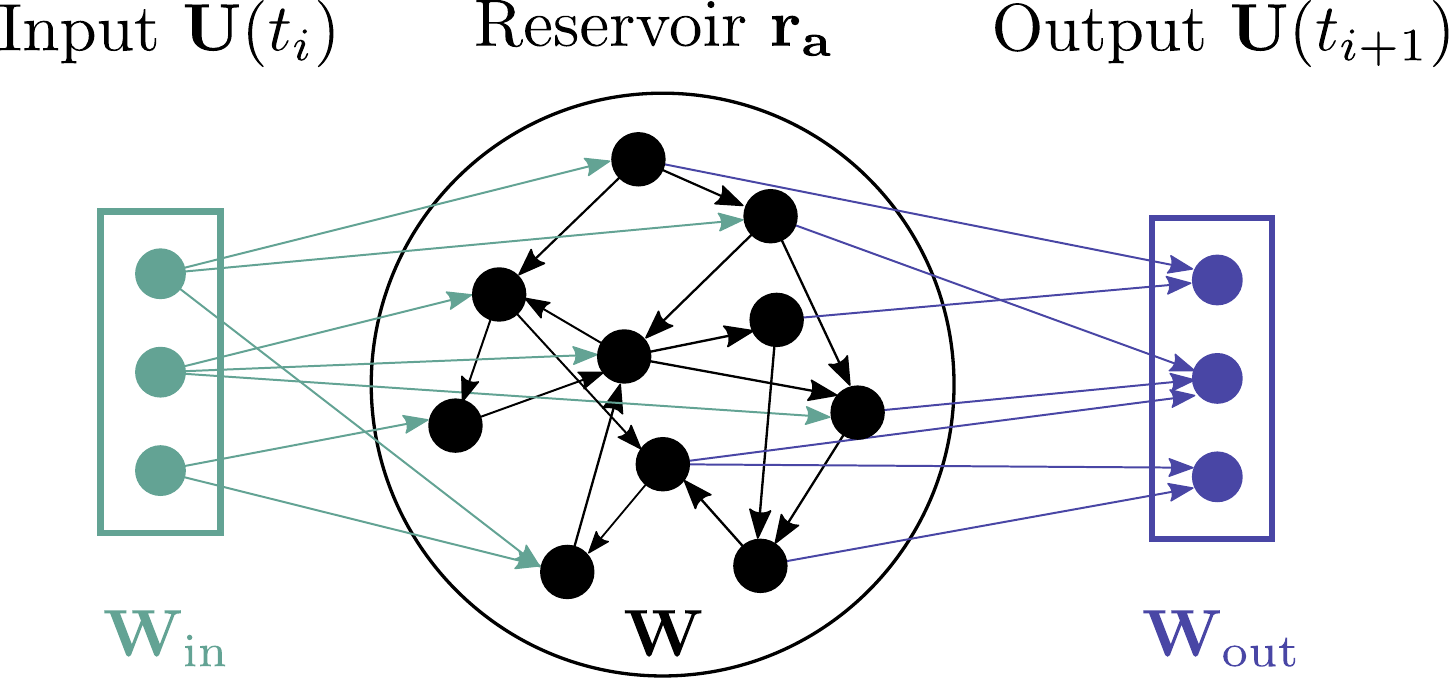}
    \caption{Schematic representation of an echo state network.}
    \label{fig:ESN}
\end{figure}
At every time $t_i$, the input bias $\vect{U}(t_i)$ and the state of the reservoir at the previous time step $\vect{r}(t_{i-1})$ are combined to predict the reservoir state at the current time as well as the bias at the next time step $\vect{U}(t_{i+1})$ such that
\begin{equation}
\label{eq:state_step}
        \textbf{r}(t_i) = \textrm{tanh}\left(\mathbf{W}_{\mathrm{in}}[\mathbf{\Tilde{\mathbf{U}}}(t_i);0.1]+
        \mathbf{W}\textbf{r}(t_{i-1})\right)\quad \mtxt{and} \quad \mathbf{U}(t_{i+1}) = \mathbf{W}_{\mathrm{out}}[\mathbf{r}(t_i);1]
\end{equation}
where $\mathbf{\Tilde{\mathbf{U}}}$ is the input bias normalised by the range component--wise; and the constant $0.1$ is used to break the symmetry of the ESN~\citep{huhn_learning_2020}. The operator $[\,;\,]$ indicates vertical concatenation. The matrices 
$\mathbf{W}_{\mathrm{in}}$ and $\mathbf{W}$ are predefined as fixed, sparse and randomly generated. Specifically, $\mathbf{W}_{\mathrm{in}}$ has only one non-zero element per row, which is sampled from a uniform distribution in $[-\sigma_{\mathrm{in}},\sigma_{\mathrm{in}}]$, where $\sigma_{\mathrm{in}}$ is the input scaling. $\textbf{W}$ is an Erdős--Renyi matrix with average connectivity $d=5$, in which each neuron (each row of $\mathbf{W}$) has on average $d$ connections (non--zero elements), which are obtained by sampling from a uniform distribution in $[-1,1]$; the entire matrix is then re--scaled by a multiplication factor to set the spectral radius, $\rho$. The weights of $\mathbf{W}_{\mathrm{out}}$ are determined through training, which consists of solving a linear system for a training set of length $N_\mtxt{tr}$
\begin{equation}
\label{eq:RidgeReg}
    (\mathbf{R}\mathbf{R}^\mtxt{T} + \gamma_t \mathbb{I})\mathbf{W}_{\mathrm{out}}^\mtxt{T} = \mathbf{R} \mathbf{U}_{\mathrm{train}}^\mtxt{T},
\end{equation}
where $\mathbf{R}\in\mathbb{R}^{(N_{r}+1)\times N_{\mathrm{tr}}}$ is the horizontal concatenation of the augmented reservoir state for each time in the training set, $[\mathbf{r}(t_i);1]$ with $i=1,...,N_\mtxt{tr}$; $\mathbf{U}_{\mathrm{train}}\in\mathbb{R}^{N_\mtxt{mic}\times N_{\mathrm{tr}}}$ is the time--concatenation of the concatenation of the output data; 
$\mathbb{I}$ is the identity matrix; and $\gamma_t$ is the Tikhonov regularisation parameter \citep{lukovsevivcius_practical_2012}.  
We employ Recycle Validation \citep{racca_robust_2021} to select the input scaling, $\sigma_\mathrm{in}=0.0126$, the spectral radius, $\rho=0.9667$, and Tikhonov parameter, $\gamma_t=1\E{-16}$. 

\subsection{Thermoacoustic Echo State Network}
\begin{figure}
    \centering
    \includegraphics[width=\linewidth]{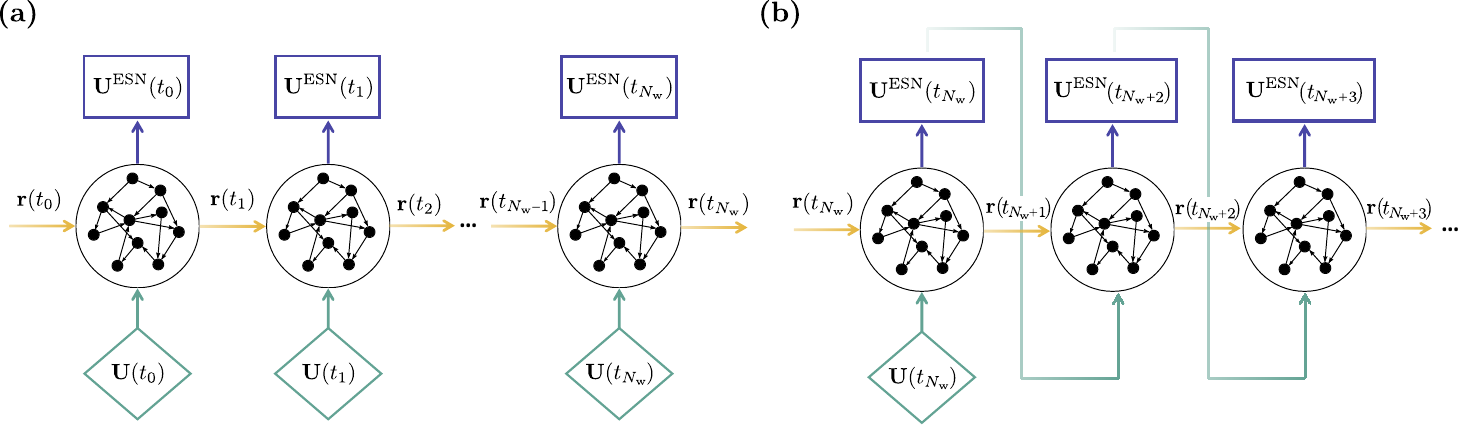}
    \caption{Schematic representation of the echo state network forecast methods. (a) Open loop forecast, and (b) closed loop estimation of the bias.}
    \label{fig:ESN_unfolded}
\end{figure}

On the one hand, in open loop (Fig.~\ref{fig:ESN_unfolded}{a}), the true bias data~\eqref{eq:bias} is fed into every forecast step to compute \eqref{eq:state_step}. The  output bias from the ESN is disregarded in open loop, which is used to initialise the network (\textit{the washout}), such that $\vect{U}_\mtxt{w}\in\mathbb{R}^{N_\mtxt{mic}\times N_\mtxt{w}}$. 
State and parameters are not updated during the washout. 
On the other hand, in closed loop  (Fig.~\ref{fig:ESN_unfolded}{b}), the true bias data~\eqref{eq:bias} is fed in the first step, then the output bias from a forecast step \eqref{eq:state_step} is used as the initial condition of the next step. The ESN forecast frequency is set to be five times smaller than the thermoacoustic model time step, to reduce the additional computation cost associated with the bias estimation. 
In detail, the pseudo--algorithm~\ref{alg1} summarises the procedure we propose for bias--aware data assimilation with an echo state network: 
(1) the ensemble of acoustic modes is initialised and forecast for the washout time; 
(2) we run the ESN in open loop to initialise the reservoir; and  
(3) we perform data assimilation. 
When measurements become available, we compute the  ensemble of pressures by adding the estimated bias from the ESN, {$\mathbf{U}$}, to the expectation of the forecast pressure at the microphones, such that the unbiased ensemble is centred around the unbiased pressure state  
$
    \vect{\hat{p}'_{\mathrm{mic\,}_j}} =  \vect{U} + \vect{{p}'_{\mathrm{mic\,}_j}} \; \mtxt{for}\; j = 1,...,m
$, 
where the $(\,\hat{\,}\,)$ indicates a statistically--unbiased quantity. 
Subsequently, we perform the analysis step. 
The EnSRKF obtains the optimal combination between the unbiased pressures and the observations, which indirectly updates the biased ensemble of acoustic modes and parameters. 
The resulting analysis ensemble is used to re--initialise the ESN for the next forecast in closed--loop, such that the bias is the difference between the observations and the expectation of the analysis pressures, i.e., $\vect{U} = \vect{{p}'^{~t}_{\mathrm{mic}}} - \left\langle\vect{{p}'^{~a}_{\mathrm{mic}}}\right\rangle$. 
Finally, the Rijke model and the ESN are time--marched until the next measurement becomes available for assimilation. 

\begin{algorithm}
\caption{Data assimilation with ESN bias estimation}
\label{alg1}
\begin{algorithmic}
\State $A^\mtxt{f}(t_0)\gets $ \Call{InitialiseEnsemble}{}
\While{$t\leq t_\mtxt{washout}$}\Comment{Get ESN running}
\State ${p^\mtxt{f}(t)}\gets $ \Call{ComputePressure}{$A^\mtxt{f}(t)$}~\eqref{eq:microphones}
\State $U(t) = p^\mtxt{true} - <p^\mtxt{f}>$ 
\State $A^\mtxt{f}(t+\Delta t) \gets $  \Call{ForecastRijkeModel}{$A^\mtxt{f}(t)$}~\eqref{eq:ivp_compact}
\State $t = t + \Delta t$
\EndWhile
\State $[U(t), r(t)] \gets $ \Call{OpenLoopESN}{$U(t_0:t_\mtxt{washout})$}~\eqref{eq:state_step}
\Loop{: for every $t$ when observations  $p^\mtxt{true}$ are available}\Comment{Start data assimilation}
\State ${p^\mtxt{f}(t)}\gets $ \Call{ComputePressure}{$A^\mtxt{f}(t)$}
\State ${\hat{p}^\mtxt{f}_j(t)}= {p^\mtxt{f}_j(t)} +\,U(t)$, for $j=1:m$
\State $A^\star \gets $\Call{AugmentState}{$A^\mtxt{f},{\hat{p}^\mtxt{f}} $}
\State $A^\mtxt{a} \gets $ \Call{EnSRKF}{$A^\star, p^\mtxt{true}$}~\eqref{th:EnSRKF}
\State ${p^\mtxt{a}(t)}\gets $ \Call{ComputePressure}{$A^\mtxt{a}(t)$}
\State $U(t) = p^\mtxt{true} - <p^\mtxt{a}>$ 
\While {no observations are available at time $t$}
\State $[U(t+\Delta t),r(t+\Delta t)] \gets $  \Call{ClosedLoopESN}{$U(t), r(t)$}~\eqref{eq:state_step}
\State $A^\mtxt{f}(t+\Delta t) \gets $  \Call{ForecastRijkeModel}{$A^\mtxt{f}(t)$}
\State $t = t + \Delta t$
\EndWhile
\EndLoop
\end{algorithmic}
\end{algorithm}

\subsection{Test case}\label{sec:test_case}
The higher--fidelity models we use is based on the travelling--wave  approach of~\citet{dowling_kinematic_1999}, which is described in detail by \citet{aguilar_sensitivity_2019}. 
The  acoustic pressure and velocity are written as functions of two acoustic waves that propagate downstream and upstream of the tube~\citep[see Eqs.~(3.5) and (3.7) in][]{dowling_kinematic_1999}. As shown in Figure~\ref{fig:rijke_wave}, the waves are defined as $f$ and $g$, with convective velocities $\tilde{c}_{0}\pm\tilde{u}_{0,\mtxt{u}}$ in the region $\tilde{x} \leq \tilde{x}_\mtxt{f}$; and $h$ and $j$, with convective velocities $\tilde{c}_{0}\pm\tilde{u}_{0,\mtxt{d}}$ in $\tilde{x} \geq \tilde{x}_\mtxt{f}$. This model uses dimensional quantities, so we transform our dimensionless thermoacoustic system \eqref{eq:final_syst} into its dimensional form. The equations that relate the travelling waves $g(\tilde{t})$ and $h(\tilde{t})$ to the heat release {rate} $\tilde{\dot{Q}}(\tilde{t})$ are, with a slight abuse of notation, 
\begin{equation}\label{eq:wave}
    \vect{\mathcal{X}} \begin{pmatrix}
    g(\tilde{t})\\
    h(\tilde{t})
    \end{pmatrix} =    \vect{\mathcal{Y}} \begin{pmatrix}
    g(\tilde{t}-\tilde{\tau}_\mtxt{u})\\
    h(\tilde{t}-\tilde{\tau}_\mtxt{d}))
    \end{pmatrix} + \begin{pmatrix}
    0\\
    \left(\tilde{\dot{Q}}(t)-\tilde{\dot{Q}}_0\right)/\left(A\tilde{c}_{0,\mtxt{u}}\right)
    \end{pmatrix}
\end{equation}
where $A$ is the cross--sectional area of the duct;   $\tilde{\tau}_\mtxt{u}$ and $\tilde{\tau}_\mtxt{d}$ are the times taken for the acoustic waves to travel from the flame to the upstream and downstream boundaries, respectively. 
For completeness, the matrices $\vect{\mathcal{X}}$ and $\vect{\mathcal{Y}}$ are provided in the Supplementary material. 
 
\begin{figure}
    \centering
    \includegraphics[width=.9\textwidth]{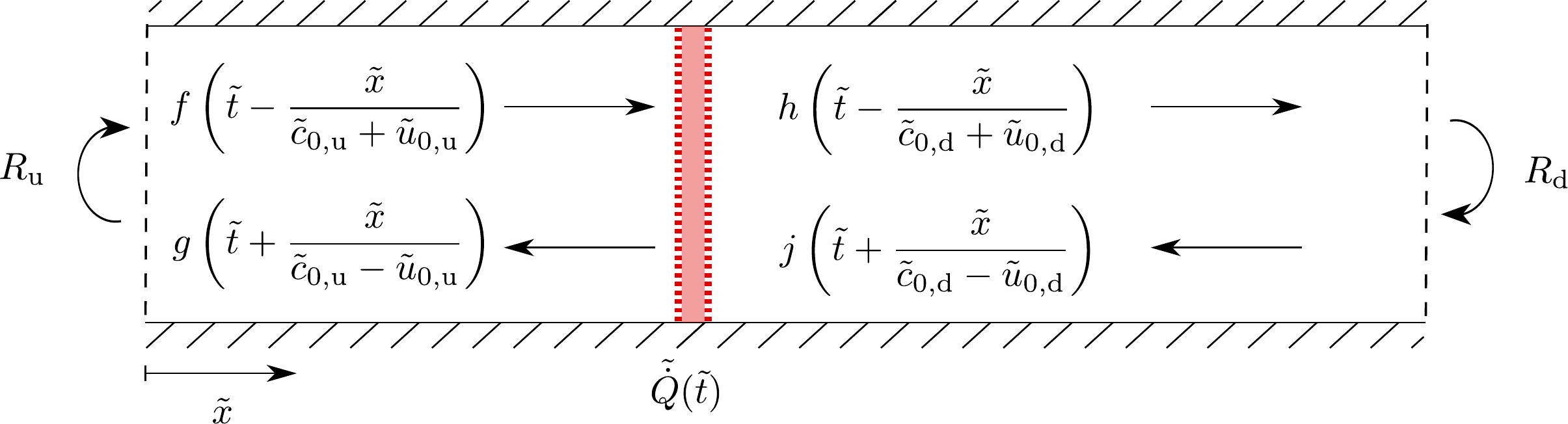}
    \caption{Schematic of the acoustic waves travelling upstream and downstream of a flame in an open--ended tube.}
    \label{fig:rijke_wave}
\end{figure}

In contrast to the low--fidelity model of~\S~\ref{sec:Rijke}, the travelling--wave approach 
(i) accounts for mean flow effects; and 
(ii) accounts for non--ideal open boundary conditions, such that $f(\tilde{t}) = R_\mtxt{u}\,g(\tilde{t}-\tilde{\tau}_\mtxt{u})$ and $j(\tilde{t}) = R_\mtxt{d}\,h(\tilde{t}-\tilde{\tau}_\mtxt{d})$, where $R_\mtxt{u}$ and $R_\mtxt{d}$ are the reflection coefficients. We set the mean flow velocity upstream of the flame to $\tilde{u}_{0,\mathrm{u}}=10~\mtxt{m/s}$, and the mean heat release rate to $\bar{\dot{Q}}_0=2000~\mtxt{W}$. The velocity downstream of the flame is computed by applying the jump conditions in the energy and momentum equations (see (3.2) and (3.3) in \citep{dowling_kinematic_1999}). We set dissipative boundary condition to $R_\mtxt{u}=R_\mtxt{d}=-0.99$. 
In contrast to the simple heat--release law of the low--fidelity model of~\S~\ref{sec:Rijke}, the heat release rate is computed with a flame kinematic model, which kinematically tracks the flame front area~\citep[detailed code in][]{aguilar_sensitivity_2019}. 

We test the bias--aware data assimilation with the echo state network for a limit cycle. We perform unbiased state estimation, and combined unbiased state and parameter estimation using synthetic pressure data obtained from the travelling--wave and  flame kinematic model as the truth (higher--fidelity data). 
For brevity, we perform parameter estimation to the heat source strength $\tilde{\beta}$ only, which we physically expect to be  $\tilde{\beta}\sim\mathcal{O}(10^6)~\mtxt{\frac{W s^{1/2}}{m^{5/2}}}$~\citep[Eq.~(2.7) in][]{juniper2011}. 
The results are discussed in Section~\ref{sec:bias_results}.
}

\section{Nonlinear characterisation}\label{sec:characterisation}

In order to assess the performance of data assimilation, we first  characterise the nonlinear  dynamics by analysing the solutions at regime (after the initial transient) with bifurcation analysis and  nonlinear time series analysis~\citep{kantz_nonlinear_2003,kabiraj_bifurcations_2011,guan2020intermittency}.  
 The system's  parameters are $x_\mtxt{f} = 0.2$, $C_1 = 0.1$, $C_2=0.06$ and $N_m= 10$. 

\begin{figure}
    \centering
    \includegraphics[width=.6\textwidth]{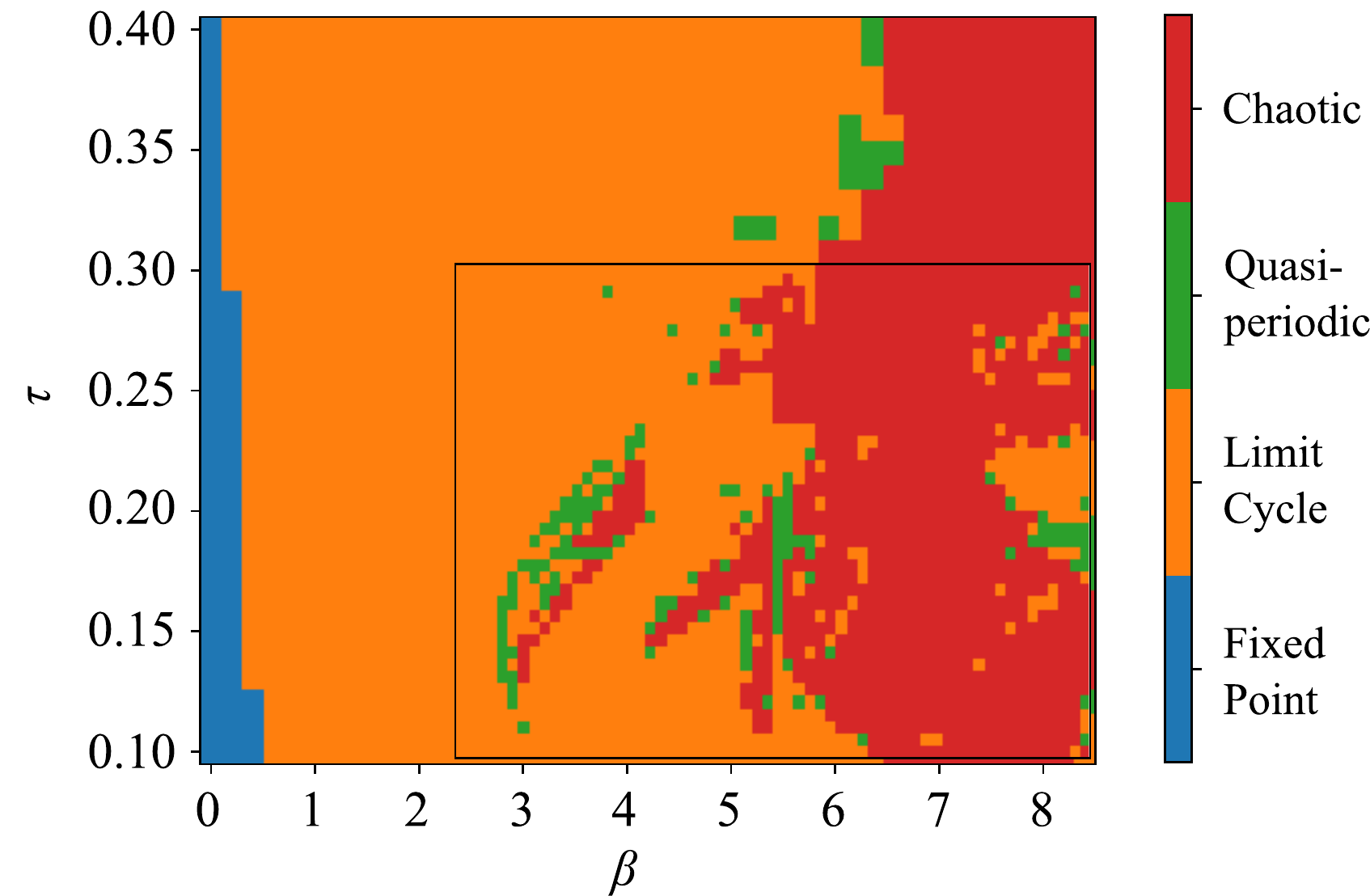}
    \caption{Two--dimensional bifurcation diagram. Classification of the attractor of the thermoacoustic system. The area enclosed by the black rectangle corresponds to a refined grid. The coarse and fine sweeps are performed with resolutions $(\Delta\beta, \Delta\tau)= (0.2, 0.01)$ and $(\Delta\beta, \Delta\tau)= (0.1, 0.005)$, respectively.
    }
    \label{fig:2D_bif_diagram}
\end{figure}

In bifurcation analysis, we examine the topological changes in the pressure oscillations, $p_\mtxt{f}'$, as the control parameters vary. First, we study the two--dimensional bifurcation diagram, which is shown in
Figure~\ref{fig:2D_bif_diagram}. 
The classification in the two--dimensional diagram is obtained following the procedure of \citet{huhn_stability_2019}. 
This method consists of obtaining the Lyapunov exponents, $\lambda_i$ through covariant--vector analysis. 
With this, the dynamical motions are identified as: 
(i) fixed point if $\lambda_1 < 0$; 
(ii) limit cycle if $\lambda_1 = 0$ and $\lambda_2 < 0$; 
(iii) quasiperiodic if $\lambda_1 = 0$ and $\lambda_2 = 0$; 
and (iv) chaotic if $\lambda_1 > 0$. 
For small $\beta$ and $\tau$ the system converges to a fixed point because the thermoacoustic energy is smaller than damping. 
As the heat source strength increases, the Rayleigh criterion is fulfilled and self--excited oscillations arise as limit cycles. When $\beta$ reaches values over 2.5, different types of solution appear, such as
quasiperiodic or chaotic attractors.  
The refined region in Figure~\ref{fig:2D_bif_diagram} shows that the type of solutions is sensitive to small changes in the control parameters, which has implications for data assimilation, as argued in the remainder of the paper. 

\begin{figure}
    \centering
  \includegraphics[width = \textwidth ]{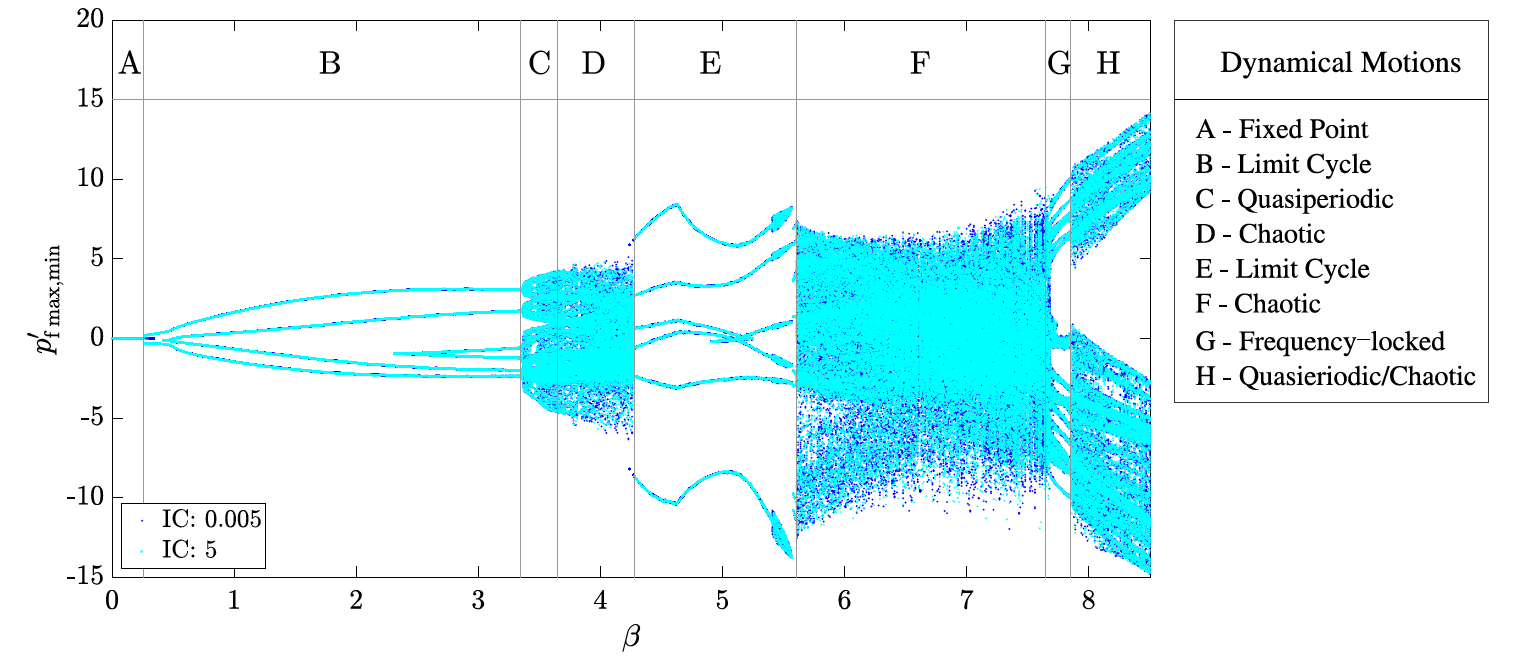}
    \caption{One--dimensional bifurcation diagram. Maxima and minima of the pressure oscillations at the flame location versus the heat--source strength. The solutions obtained for small/large initial conditions are shown in a dark/light blue colour.  This diagram identifies different nonlinear behaviours, which have implications for data assimilation.
    }
    \label{fig:bifurcation_diagram}
\end{figure}
  
These topological changes are further investigated with a one--dimensional bifurcation diagram for a fixed time delay ($\tau=0.2$), shown in Figure~\ref{fig:bifurcation_diagram}.
Because the nonlinear solutions at regime may vary with the initial condition, two sets of results are shown for a small initial condition ($\eta_j= \dot{\eta}_j /j \pi= 0.005$) and a large initial condition ($\eta_j= \dot{\eta}_j /j \pi= 5$) to capture subcritical behaviours. 
The bifurcation diagram is obtained by marching forward in time the governing  equations of the nonlinear dynamical system until the system reaches a statistically stationary state. 
For each value of the control parameter, the bifurcation  diagram shows the peaks and troughs of the acoustic pressure at the flame location. (The nonlinear time series analysis results are shown in the {Supplementary material}.)

From left to right, first, 
 the solution is the fixed point (region~A), which is the case of no oscillations. 
Second, the appearance of periodic oscillations from a fixed point is observed with a large initial condition at $\beta = 0.26$, with a small region of hysteresis from $\beta = 0.26$ to $\beta = 0.34$. 
This first self--sustained state is a period--1 limit cycle (region~B), which originates from a subcritical Hopf bifurcation. 
Within region B, the system undergoes a period--doubling bifurcation at $\beta = 0.6$ from period--1 to period--2 oscillations. 
%
Third, the period--2 limit cycle bifurcates into a 3--torus quasiperiodic motion at $\beta = 3.35$ (region~C). 
A quasiperiodic oscillation is an aperiodic solution that results from the interaction between two or more incommensurate frequencies, (also known as a Neimark--Sacker bifurcation)~\citep{kabiraj_bifurcations_2011}.
%
Fourth, the solution becomes chaotic at $\beta=3.65$ (region~D). 
In summary, the evolution from region~A to region~D shows that the system reaches a chaotic state via a quasiperiodic route to chaos, i.e., via a Ruelle--Takens scenario~\citep{kabiraj_route_2012}. 
Fifth, after this first route to chaos, changes in the control parameter drive the system back to a periodic limit cycle through a tangent bifurcation~\citep{kantz_nonlinear_2003} at approximately $\beta = 4.25$  (region~E), with a second region of hysteresis from $\beta = 4.24$ to $\beta = 4.28$. 
These high amplitude limit cycles region  becomes again  chaotic at $\beta=5.61$ (region~F). 
Sixth, when $\beta$ reaches $7.65$, the system evolves towards a frequency--locked state (region~G).  
Frequency--locked solutions arise from the competition between two or more frequencies, but in contrast to quasiperiodic signals, these frequencies are commensurate. 
%
Seventh, at $\beta = 7.85$, the frequency--locked solution bifurcates into a quasiperiodic solution (region~H). 
Region--H solutions show a two--dimensional toroidal structure, in contrast to the three--dimensional torus from region~C. 
In region~H, some of the simulations showed that there are areas of chaotic dynamics, which can be appreciated by the difference of the solutions from the small and large initial condition in Figure~\ref{fig:bifurcation_diagram}. 
(A higher region refinement could be performed to fully understand the bifurcations within this region, however, this is beyond the scope of this work.)
The qualitative bifurcation behaviour of this reduced--order model is observed in experiments~\citep{kabiraj_bifurcations_2011,kabiraj_nonlinear_2012}, which means that the reduced--order model qualitatively captures the nonlinear thermoacoustic dynamics.  

The bifurcation analysis shows a rich variety of solutions in a relatively small range of parameters, i.e., small changes of a parameter, or a state, can generate solutions that are topologically different.
This nonlinear sensitivity has implications in the design of a data--assimilation ensemble framework, as discussed in~\S~\ref{sec:results}. 

\section{Twin experiments in non--chaotic regimes}\label{sec:results}

We perform a series of experiments with synthetic data, which is generated by the model. 
To mimic an experiment, we add stochastic uncertainty to the synthetic data by prescribing an observation covariance matrix. 
This approach is also known as the twin experiment~\citep[e.g., ][]{traverso_data_2019}.
The EnSRKF algorithm is tested in the different regions of Figure~\ref{fig:bifurcation_diagram}, 
for the different nonlinear regimes: 
fixed point, limit cycle, frequency--locked, quasiperiodic and chaotic. 
The filter is first tested in the non--chaotic regimes for the assimilation of 
(i) acoustic modes (\S~\ref{sec:DA_modes}), 
 and
 (ii)  acoustic pressure from microphones (\S~\ref{sec:DA_mics}). 
The assimilation of chaotic solutions, which presents further challenges, is investigated in~\S~\ref{sec:CH_results}. 
Different simulations are performed to determine suitable values for the number of ensemble members ($m$); 
the time between analysis ($\Delta t\mtxt{_{analysis}}$); 
the standard deviation ($\sigma_\mtxt{frac}$), i.e., the observations' uncertainties during the acoustic modes assimilation; 
and the standard deviation of the microphone measurements ($\sigma_\mtxt{mic}$).  
Table~\ref{tab:KF_simulation_parameters} shows the parameters and initial conditions of the reference (i.e., ``true'') solution.
This range of parameters is justified from the literature in thermoacoustic data assimilation~\citep{traverso_data_2019}.
Computational time is discussed in the {Supplementary material}. 

\begin{table}
\centering
\caption{Parameters and initial conditions for the true solution.}
\label{tab:KF_simulation_parameters}
\begin{tabular}{@{}llllll@{}}
\toprule
Parameter & Value & $ \qquad$ & & Parameter & Value \\ \midrule
$x_\mtxt{f}$ & 0.2 &  & & $\beta$ & [0.2, 0.4, 3.6, 7.7, 7.0] \\
$N_m$ & 10 &  & & $\tau$ & 0.2 \\
$N_c$ & 10 &  & &$m$ & 10\\
$\Delta\, t$ & 0.001 &  & & $\sigma_\mtxt{frac} $ & 0.25   \\
$\eta_j (t=0)$ & 0.005 &  & & $\Delta t\mtxt{_{analysis}}$ &[$2$ (Non--chaotic), $0.5$ (Chaotic)]\\
$\dot{\eta}_j /j \pi (t=0)$ & 0.005 &  & &    $N_\mtxt{mic}$ & 6 \\
$v_i(t=0)$ & 0 &  & & $\sigma_\mtxt{mic}$ & 0.01  \\ \bottomrule
\end{tabular}
\end{table}

\subsection{Assimilation of the acoustic modes}\label{sec:DA_modes}
This section includes results for state estimation (\S~\ref{sec:modes_SE}) and combined state and parameter estimation (\S~\ref{sec:modes_SPE}).
\subsubsection{State estimation}\label{sec:modes_SE}
\begin{figure} 
  \centering
        \includegraphics[width=\textwidth]{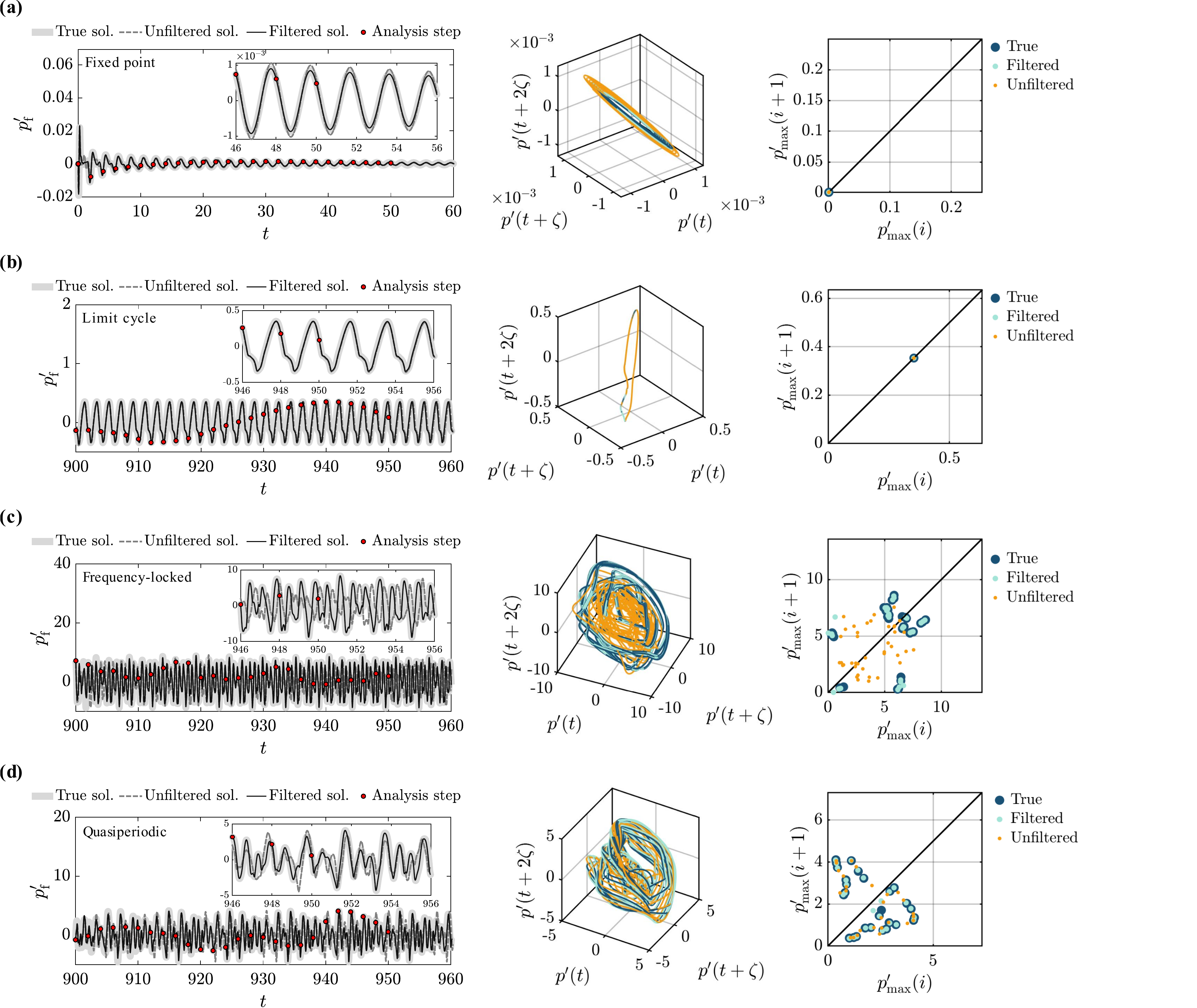}
  \caption{
    {Real--time learning of the state. Assimilation of acoustic modes for state estimation of non--chaotic regimes. 
  (a) Transient towards a fixed point ($\beta=0.2$); 
  (b) limit cycle ($\beta=0.4$); 
  (c) frequency--locked ($\beta=7.7$); 
  and 
  (d) quasiperiodic ($\beta=3.6$). 
  Left: True pressure oscillations at the flame location (light grey), unfiltered solution (dashed dark grey) and  filtered solution (black). The analysis time steps are indicated with red circles. 
  Right: Phase portrait and first return map of the true (dark blue), filtered (light blue), and unfiltered (orange) solutions.
  $m=10$, $\sigma_\mathrm{frac}=0.25$, $\Delta t\mtxt{_{analysis}}=2$. }
  }
  \label{fig:DA_modes}
\end{figure}

This section presents simulations performed assuming that there are observations available for all acoustic modes, i.e., the number of observations is $q=2N_m=20$. 
(Including observations for the {dummy velocity variables} would further improve the filter convergence, however, they are not considered because the velocity advection field in the heat source region is not measured in a real engine.)
Figure~\ref{fig:DA_modes} shows the acoustic pressure before assimilation (unfiltered solution), after assimilation (filtered solution) and the data at the assimilation steps (analysis steps).  
Panels (a) to (d) show the transient of a fixed point, a period--1 limit cycle, a frequency--locked, and a quasiperiodic motion, respectively. 
%
%
%
In the filtered solution, data assimilation is performed during the first 50 time units, and it is marched in time without further assimilation for 10 more time units. 
The EnSRKF successfully learns (i.e., infers) the true solution for all the nonlinear regimes. 
As expected, the convergence is faster for the fixed point and limit cycle cases (Figs.~\ref{fig:DA_modes}{a,b}) because they are simpler dynamical motions. 
(The unfiltered solution also converges to the same value for these simple cases. 
This is due to the stable nature of their attractors, and because their regions are unaffected by the chaotic butterfly effect.)
For multi--frequency dynamical regimes, Figures~\ref{fig:DA_modes}{c,d} show that the Bayesian update can  learn the frequency--locked and quasiperiodic states of regions C and G in Figure~\ref{fig:bifurcation_diagram}. 
However, these show more discrepancies between the filtered and true solutions. 
Physically, this is due to the multiple bifurcations that occur in a small range of parameters, which is typical of thermoacoustic systems.
In reference to Figure~\ref{fig:bifurcation_diagram}, region~C is next to the chaotic region~D; and region~G is a short range region surrounded by the chaotic region~F, and the mixed quasiperiodic--chaotic region~H. 
Therefore, the discrepancy in these cases is caused by some ensemble members falling in different basins of attraction.
To overcome this issue, we propose a strategy in~\S~\ref{sec:mics_SPE}. 

\begin{figure}
    \centering
    \includegraphics[width=\textwidth]{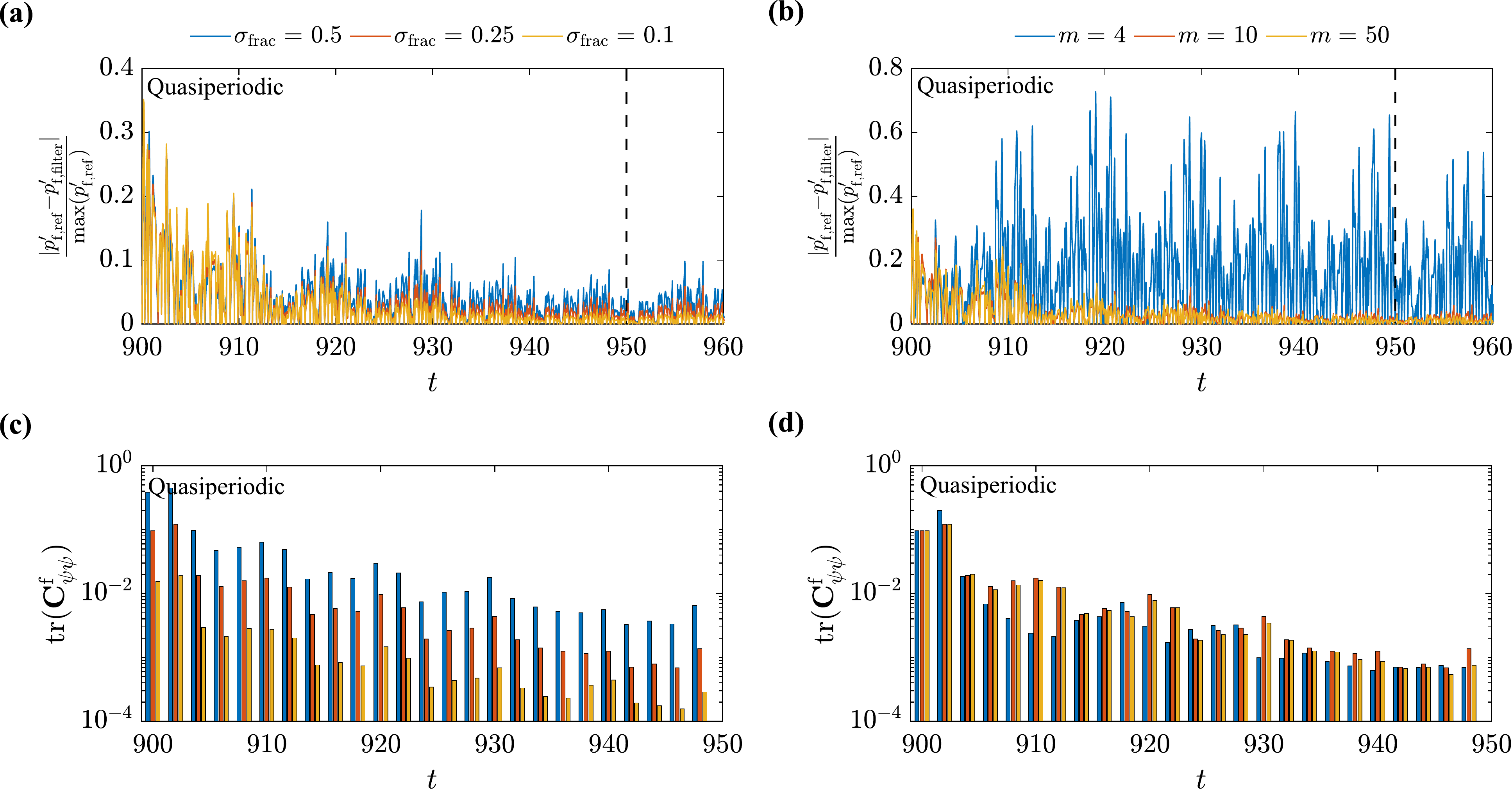}
    \caption{Assimilation of acoustic modes for state estimation of a quasiperiodic regime. Performance metrics.  
    Left: Effect of the standard deviation with $m=10$. 
    Right: effect of the ensemble size with modes measurement uncertainty $\sigma_\mathrm{frac}=0.25$. 
    The error evolution is shown with the relative difference between the filtered solutions and truth (top) and the trace of the ensemble covariance (bottom). The dashed vertical line indicates when data assimilation ends. $\beta=3.6$,  $\Delta t\mtxt{_{analysis}}={2}$.
    }
    \label{fig:DA_modes_error}
\end{figure}

The data assimilation process depends on the observation's uncertainty, $\sigma_\mtxt{frac}$, and ensemble size, $m$. 
Figure~\ref{fig:DA_modes_error} shows the performance metrics (\S~\ref{sec:method}) for the quasiperiodic solution of Figure~\ref{fig:DA_modes}{d}. 
As expected, the filtered solution is more accurate for a smaller standard deviation because the observations are closer to the truth. 
Importantly, the algorithm is capable of learning the reference solution for an ensemble having an error as large as 50\% of the mean of the acoustic modes, which means that the data assimilation algorithm is robust. 
For the pressure performance metric, the algorithm brings the relative error below 10\% after 15 time units (in the worst case scenario,  Figure~\ref{fig:DA_modes_error}{a}). 
For the covariance matrix trace performance metric, the EnSRKF continuously reduces the initial ensemble variance up to a final plateau, which cannot be zero because of the non--zero observation and forecast background noise (Figure~\ref{fig:DA_modes_error}{c}).
The evolution of the trace is an indicator of the spread of the forecast ensemble, which informs on the uncertainty of the solution. 
The  ensemble size does not have a strong influence in the ensemble uncertainty during the assimilation because the trace of the covariance matrix remains of the same magnitude independently of the value of $m$ (Figure~\ref{fig:DA_modes_error}{d}). 
Nevertheless, the relative error is significantly higher for a small ensemble with $m = 4$ (Figure~\ref{fig:DA_modes_error}{c}).
This means that four ensemble members are not sufficient to give a sufficient ensemble distribution, therefore, the solution converges to an incorrect state, but with a small spread around it. 
Comparing the errors for ten and fifty ensemble members, we see no significant differences between the solutions, which shows that having an ensemble size larger than the number of degrees of freedom is not required. 
This is one of the benefits of using the square--root filter (in the standard ensemble Kalman filter larger ensembles are needed to avoid sampling errors~\citep{livings_unbiased_2008}). 
However, the computational time required for 50 ensemble members was approximately 4 times longer than that for 10. Therefore, an ensemble size of $m=10$ provides a good approximation of the true state for the assimilation of acoustic modes, while keeping the computation time minimal. 


\subsubsection{Combined state and parameter estimation} \label{sec:modes_SPE}
\begin{figure}
    \centering
    \includegraphics[width=.99\textwidth]{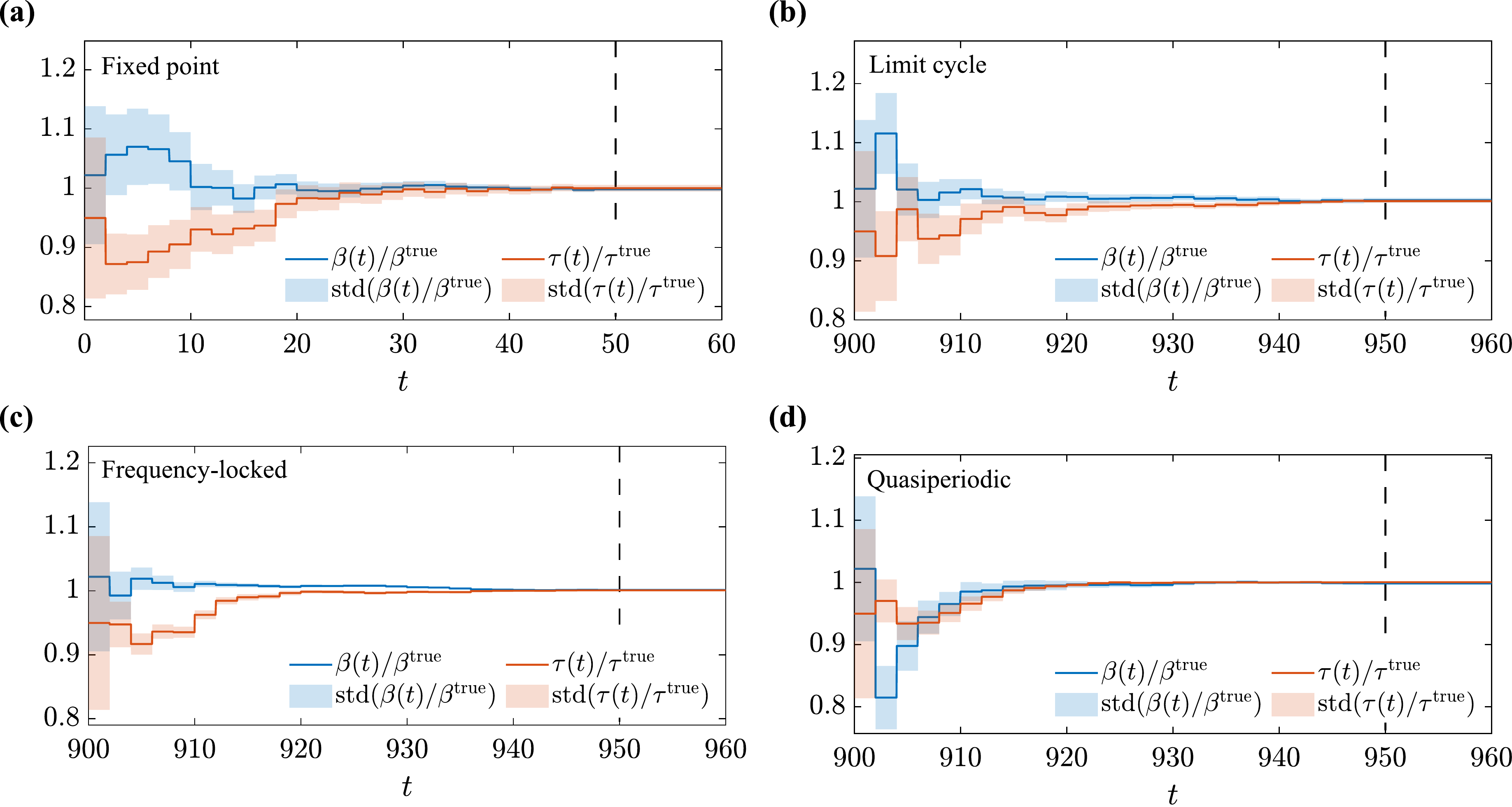}
    \caption{Real--time learning of the parameters and the state. Assimilation of acoustic modes for combined state and parameter estimation of non--chaotic regimes. 
    (a) Transient toward a fixed point ($\beta^\mtxt{true}=0.2$), 
    (b) limit cycle ($\beta^\mtxt{true}=0.4$), 
    (c) frequency--locked solution ($\beta^\mtxt{true}=7.7$), 
    and (d) quasiperiodic solution ($\beta^\mtxt{true}=3.6$). 
    The dashed vertical line indicates when data assimilation ends. $\tau^\mtxt{true}=0.2$, $m=10$, {$\sigma_\mathrm{frac}=0.25$}, $\Delta t\mtxt{_{analysis}}=2$. 
    }
    \label{fig:PE_convergence}
\end{figure}

In this section, we analyse the combined state and parameter estimation to calibrate both the state and parameters. 
The two uncertain parameters ($\vect{\alpha}=[\beta,\tau]$ in~\eqref{eq:ivp_compact_PE}) are added to the state vector and updated simultaneously with the acoustic modes and {dummy velocity variables}, as detailed in~\S~\ref{sec:Parameter_estimation}. 
Figure~\ref{fig:PE_convergence} shows the evolution of the parameters, normalised to their true value, for the four non--chaotic solutions. 
The convergence shows that the EnSRKF update is capable of learning the true $\beta$ and $\tau$ values for the four dynamical motions.

\begin{figure}
  \centering
  \includegraphics[width=.99\textwidth]{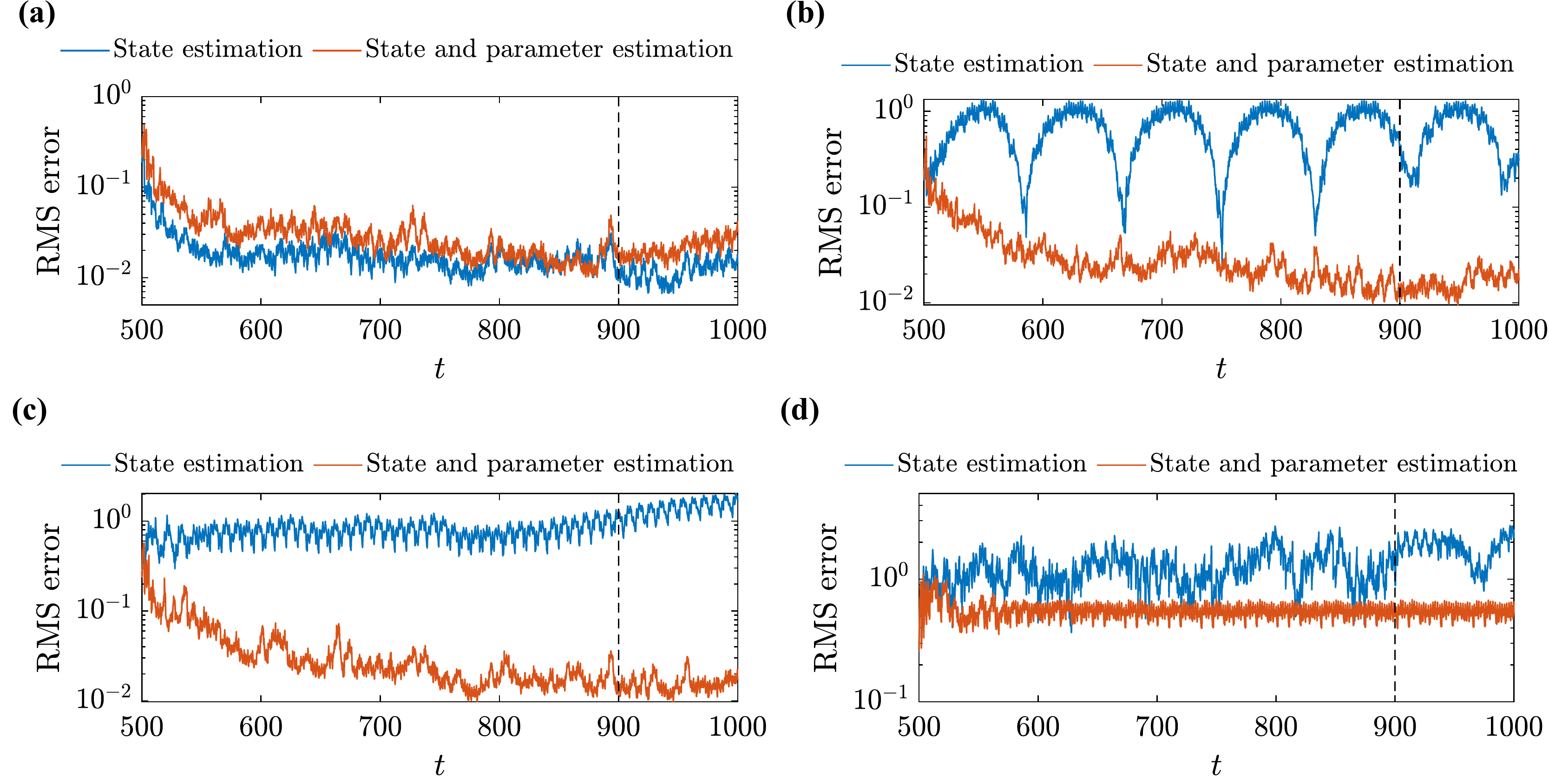}
  \caption{Assimilation of acoustic modes of a quasiperiodic regime. Performance of state estimation (blue) vs. combined state and parameter estimation (orange) in a quasiperiodic regime.
   Initial conditions $\beta= c\,\beta^\mtxt{\,true}$ and $\tau=c\,\tau^\mtxt{\,true}$ with (a) $c=1$, (b) $c=1.05$, (c) $c=1.25$, (d) $c=1.5$; and $\beta^\mtxt{true}=3.6$, $\tau^\mtxt{true}=0.2$. 
    The dashed vertical line indicates when data assimilation ends.}
  \label{fig:PE_RMS_error}
\end{figure}

For a comparison of combined state and parameter estimation with state estimation, we compute the root mean square (RMS) error. 
The RMS error at each time step is defined as the square--root of the trace of the covariance matrix of the filtered ensemble, relative to the true solution 
\begin{equation}
    \label{eq:RMS_error}
    \mtxt{RMS\,error} = \sqrt{\mtxt{tr}\left(\dfrac{1}{m - 1}\sum_{j = 1}^{m}{\left(\vect{\psi}_j-{\vect{\psi}}^\mtxt{true}\right) \,\left(\vect{\psi}_j-{\vect{\psi}}^\mtxt{true}\right)^\mtxt{T}}\right)}
\end{equation}
The RMS error is evaluated for the state estimation and the combined state and parameter estimation cases, using different initial uncertainties for $\beta$ and $\tau$. 
This is achieved in state estimation by defining $\beta= c\,\beta^\mtxt{\,true}$ and $\tau=c\,\tau^\mtxt{\,true}$, where $c$ is the defined initial uncertainty. 
For the combined state and parameter estimation, the initial $\beta$ and $\tau$ of each member in the ensemble are taken from an uniform distribution centred around $c\,\beta^\mtxt{\,true}$ and $c\,\tau^\mtxt{\,true}$, with a sample standard deviation of 25\%. 
Figure~\ref{fig:PE_RMS_error}{a} shows the RMS error for the initial parameters set to their true value. 
The state estimation only outperforms the combined state and parameter estimation in this case, as the state estimation model works with constant true parameters while the combined state and parameter estimation updates the parameters in each analysis step with the EnSRKF update. 
The true parameters are perturbed by 5\%, 25\% and 50\%  in Figs.~\ref{fig:PE_RMS_error}{b,c,d}, respectively. 
The combined state and parameter estimation simulations are capable of learning the true state up to a 25\% error in the parameters initialisation, as the RMS error is reduced by two orders of magnitude from the initial state, such as in the case of Figure~\ref{fig:PE_RMS_error}{a}. 
Combined state and parameter estimation provides an improved approximation of the solution for the highly uncertain case of 50\% error (Figure~\ref{fig:PE_RMS_error}{d}).  

%
%
%
\subsection{Assimilation of the acoustic pressure from microphones}\label{sec:DA_mics}
As detailed in~\S~\ref{sec:state_estimation}, we consider the scenario of assimilation of pressure measurements from $N_\mathrm{mic}$ microphones, located equidistantly from the flame location. 
This section includes results for state estimation (\S~\ref{sec:mics_SE}) and combined state and parameter estimation (\S~\ref{sec:mics_SPE}).
\subsubsection{State estimation}\label{sec:mics_SE}

\begin{figure} 
  \centering
        \includegraphics[width=\textwidth]{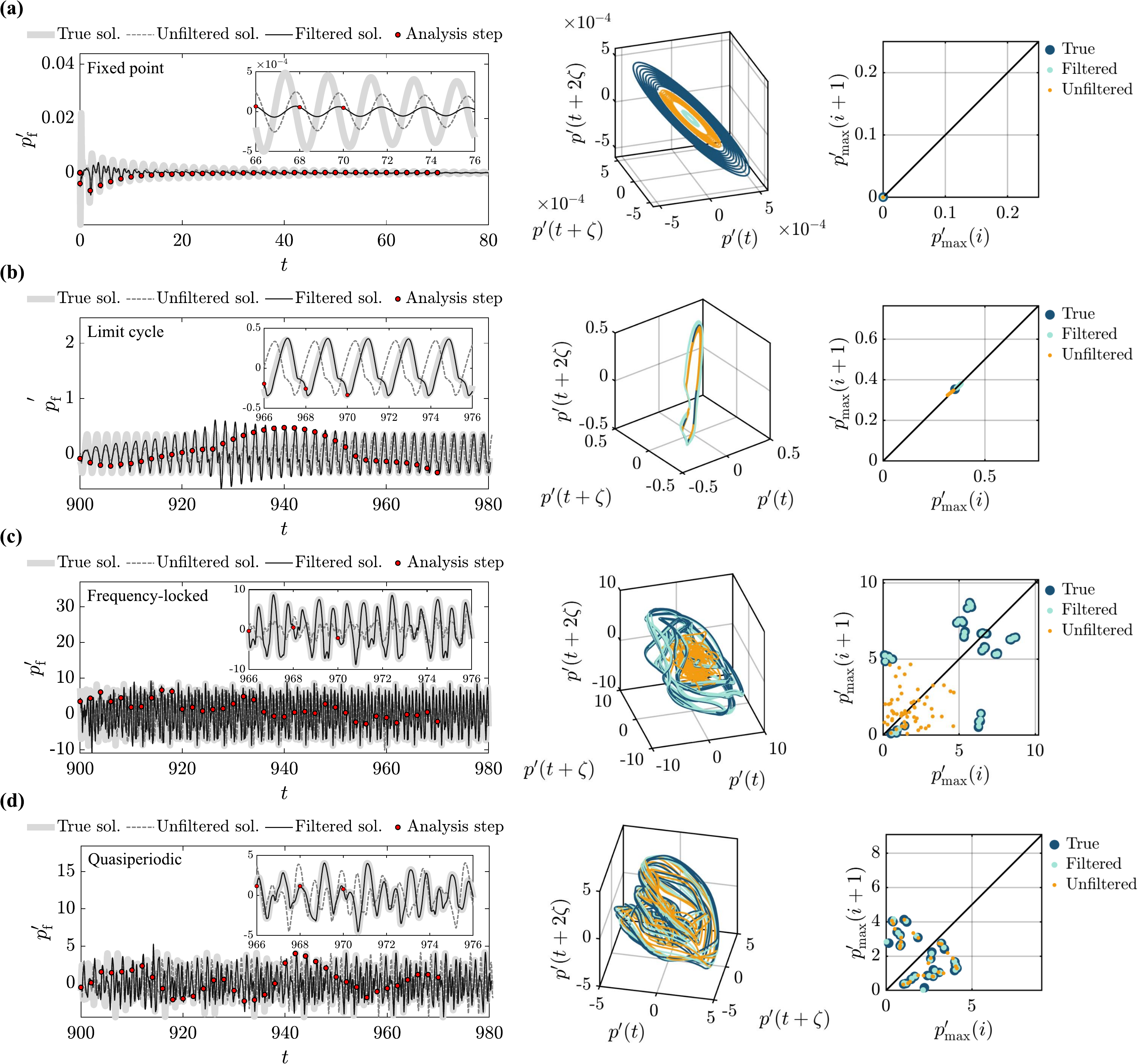}
  \caption{
    {Real--time learning of the state. Assimilation of acoustic pressure from microphones for state estimation of non--chaotic regimes.
  (a) Transient towards a fixed point ($\beta=0.2$); 
  (b) limit cycle ($\beta=0.4$); 
  (c) frequency--locked ($\beta=7.7$); 
  and 
  (d) quasiperiodic ($\beta=3.6$). 
  Left: True pressure oscillations at the flame location (light grey), unfiltered solution (dashed dark grey) and  filtered solution (black). The analysis time steps are indicated with red circles. 
  Right: Phase portrait and first return map of the true (dark blue), filtered (light blue), and unfiltered (orange) solutions.
  $m=10$,  $\sigma_\mathrm{mic}=0.01$, $\Delta t\mtxt{_{analysis}}=2$.}
  }
  \label{fig:DA_pressure}
\end{figure}

We consider a tube that is equipped with $N_\mathrm{mic}=6$ microphones, which measure multiple frequency contributions in the signal.  This value is chosen from the  literature in thermoacoustic experiments~\citep{garita_assimilation_2021}.
Figure~\ref{fig:DA_pressure} shows the acoustic pressure at the flame location of the true solution, the unfiltered solution, and the filtered solution, {as well as their phase space reconstruction and first return map (the calculation procedure can be found in the Supplementary material).} 
In nonlinear regimes, the algorithm successfully learns the  pressure state. 
The accuracy of the solution is lower than in the assimilation of the acoustic modes of~\S~\ref{sec:modes_SE} because, here, less information on the state is assimilated. 
(The filter is not designed for statistically non--stationary problems, which is why the transient fixed point solution is not fully learnt by the filter.) 

\begin{figure}
    \centering
    \includegraphics[width = \textwidth]{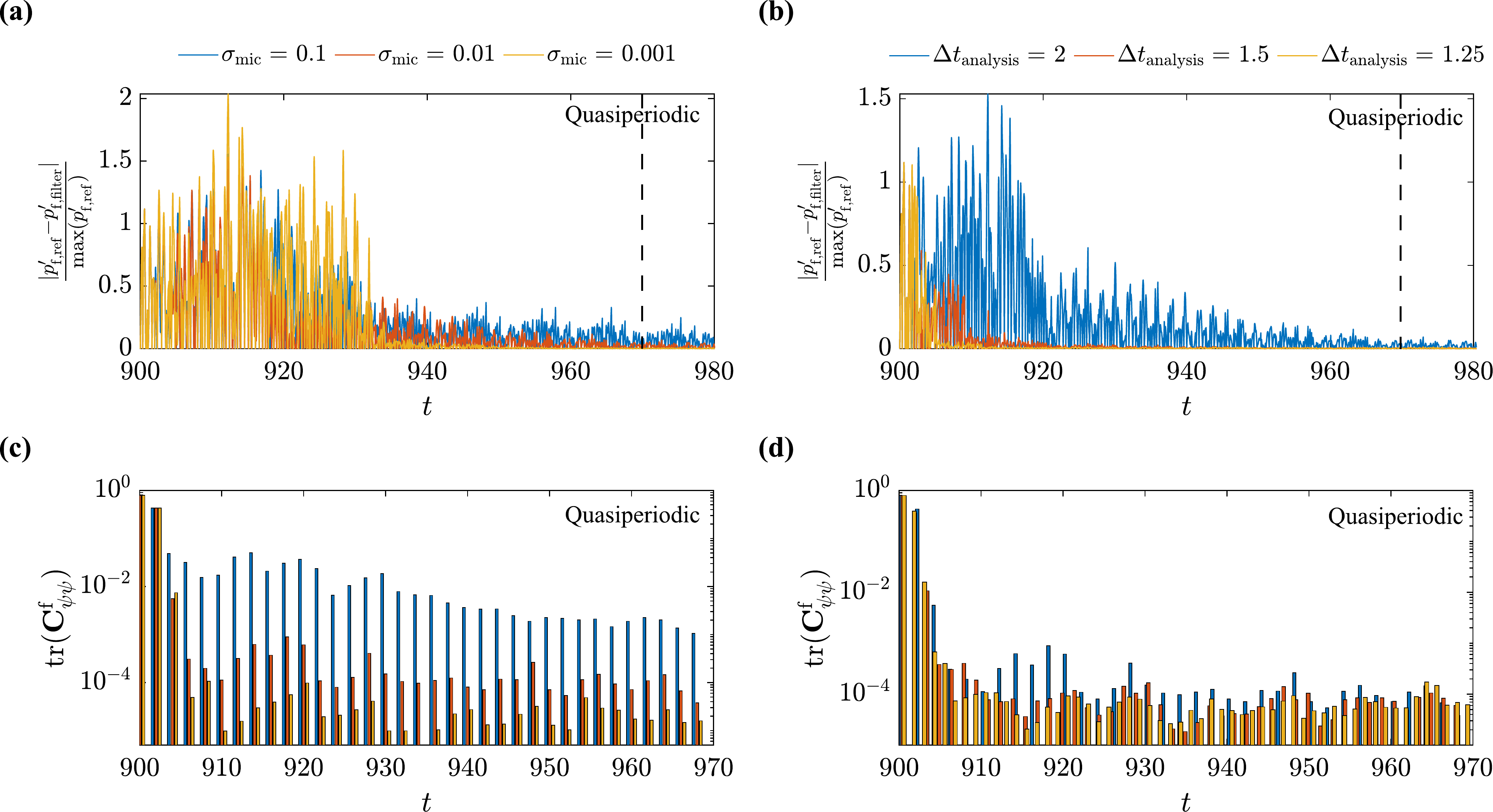}
    \caption{
    Assimilation of pressure from microphones for state estimation of a quasiperiodic regime. Performance metrics.  
    Left: Effect of the microphones' standard deviation with $\Delta t\mtxt{_{analysis}}={2}$. 
    Right: effect of the assimilation frequency with $\sigma_\mathrm{mic}=0.01$. 
    The error evolution is shown with the relative difference between the true and filtered solutions (top) and the trace of the ensemble covariance (bottom). The dashed vertical line indicates when data assimilation ends. $\beta=3.6$, $m=10$.
    }
    \label{fig:DA_pressure_error}
\end{figure}

The effect of the experimental uncertainty is analysed by varying the microphones standard deviation. 
Physically, the errors are larger than those in Figure~\ref{fig:DA_modes_error} because, here, we are  assimilating 6 components of the augmented state vector out of 36 components, whereas in \S~\ref{sec:modes_SE} the filter assimilates 20 out of the 30 components of the state vector. 
Figures~\ref{fig:DA_pressure_error}{a,c} show that, after about 20 analysis steps, the filter follows more closely the model than the observations for larger observation's uncertainties. 
(In other words, the filtered solution ``trusts'' more the prediction from the model than the observations when the experimental uncertainty is high.) 
We set {$\sigma_\mathrm{mic} = 0.01$} in the following simulations, which models experimental microphone uncertainties~\citep{de2017detection}. 
The relative error is higher than 20\%  for this case (Figure~\ref{fig:DA_pressure_error}{a}). 
Increasing the frequency of analysis allows for a faster convergence with a smaller relative error (Figs.~\ref{fig:DA_pressure_error}{b,d}). With a time between analysis of $\Delta t\mtxt{_{analysis}}=1.5$ or $1$, the relative error of the filtered solution becomes less than $10\%$ in only 10 time units, approximately. Thus, for the assimilation of microphone pressure data, a higher frequency of analysis is more suitable. 
We choose the time between analysis to~1.5 time units.
The evolution of the trace of the forecast covariance matrix indicates that the spread of the ensemble rapidly shrinks (Figs.~\ref{fig:DA_pressure_error}{c,d}). 
Besides, the spread is two orders of magnitude smaller than in the assimilation of the modes (Figs.~\ref{fig:DA_modes_error}{c,d}) and remains small even with large relative errors. 
Physically, this is because the acoustic modes are directly updated in the modes assimilation, but, in this case, the acoustic modes are unobserved variables that are updated indirectly through the microphone pressure observations.

\subsubsection{Combined state and parameter estimation} \label{sec:mics_SPE}
\begin{figure}
    \centering
    \includegraphics[width=\textwidth]{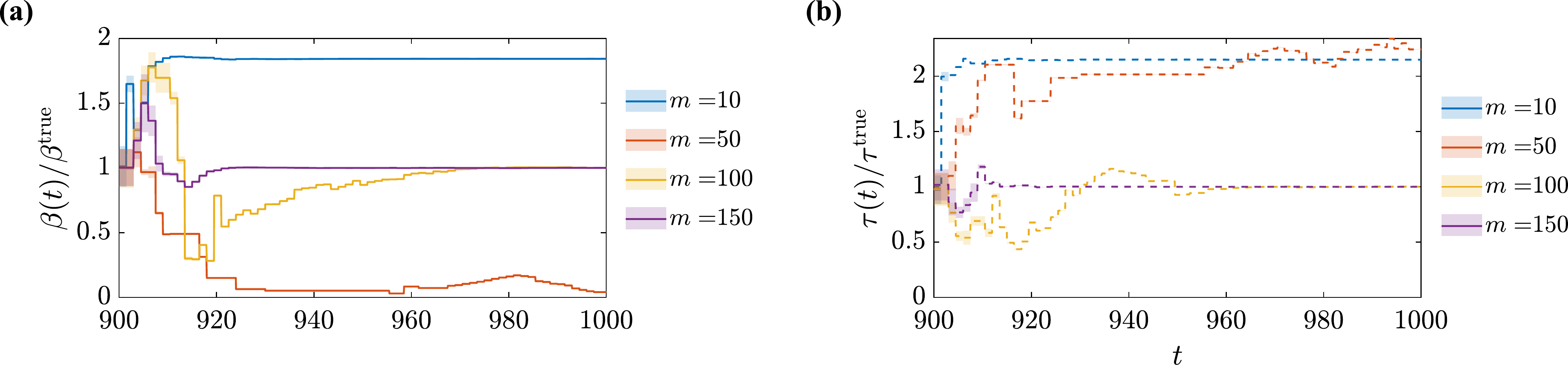}
    \caption{Real--time learning of the parameters. Assimilation of acoustic pressure from microphones for combined state and parameter estimation of a quasiperiodic solution. Left: normalised $\beta$. Right: normalised $\tau$.   $N_\mtxt{mic}=6$, $\beta^\mtxt{true}=3.6$, $\tau^\mtxt{true}=0.2$,  $\Delta t\mtxt{_{analysis}}=1.5$, $\sigma_\mathrm{mic}=0.01$. The shaded areas show the standard deviation, which becomes smaller as more data is assimilated.}
    \label{fig:increase}
\end{figure}

The parameters $\beta$ and $\tau$ are updated by the EnSRKF at each analysis step, which occurs every~1.5 time units. 
Figure~\ref{fig:increase}a,b, shows that for an ensemble of ten members, the solution converges to the parameters $\beta\approx6.6$ and $\tau\approx0.4$, which  correspond to a chaotic solution (see Figure~\ref{fig:2D_bif_diagram}). Nevertheless, the true solution is a quasiperiodic oscillator with $\beta=3.6$ and $\tau=0.2$.
This means that the filtered solution not only converges to different parameters, but also belongs to a different nonlinear regime than that of the true solution. 
Physically, this occurs because thermoacoustic dynamics experience several bifurcations in short ranges of $\beta$ and $\tau$ (Figure~\ref{fig:bifurcation_diagram}).
This makes the sampling of nonlinear thermoacoustics challenging. 
A way to circumvent this is to increase the ensemble size. %
A parametric study of the effect of the number of realisations is shown in Figure~\ref{fig:increase}.  
Ten ensemble members are not sufficient to learn the reference solution, however, the larger the ensemble, the faster the EnSRKF converges to the true solution. 

Occasionally, the EnSRKF provides  unphysical parameters as the solution of the optimisation problem, such as negative heat source strength as the solution of the optimisation problem. 
To avoid this, we reject the analysis steps that give unphysical solutions and continue the forecast with no assimilation. 
This means that we are left--truncating the Gaussian. 
Thus the parameters remain constant until the EnSRKF gives a physical solution to the optimisation problem. 
(Ad--hoc ways to bound parameters can be designed~\citep{li_constrained_2019}. This is beyond the scope of this work.) 
The thresholds for rejection are defined as $\beta \in [0.1, 10]$ and $\tau \in [0.005, 0.8]$.  
Because the rejection is effectively reducing the amount of information that can be assimilated, the ensemble convergence slows down. 
\begin{figure}
    \centering
    \includegraphics[width=\textwidth]{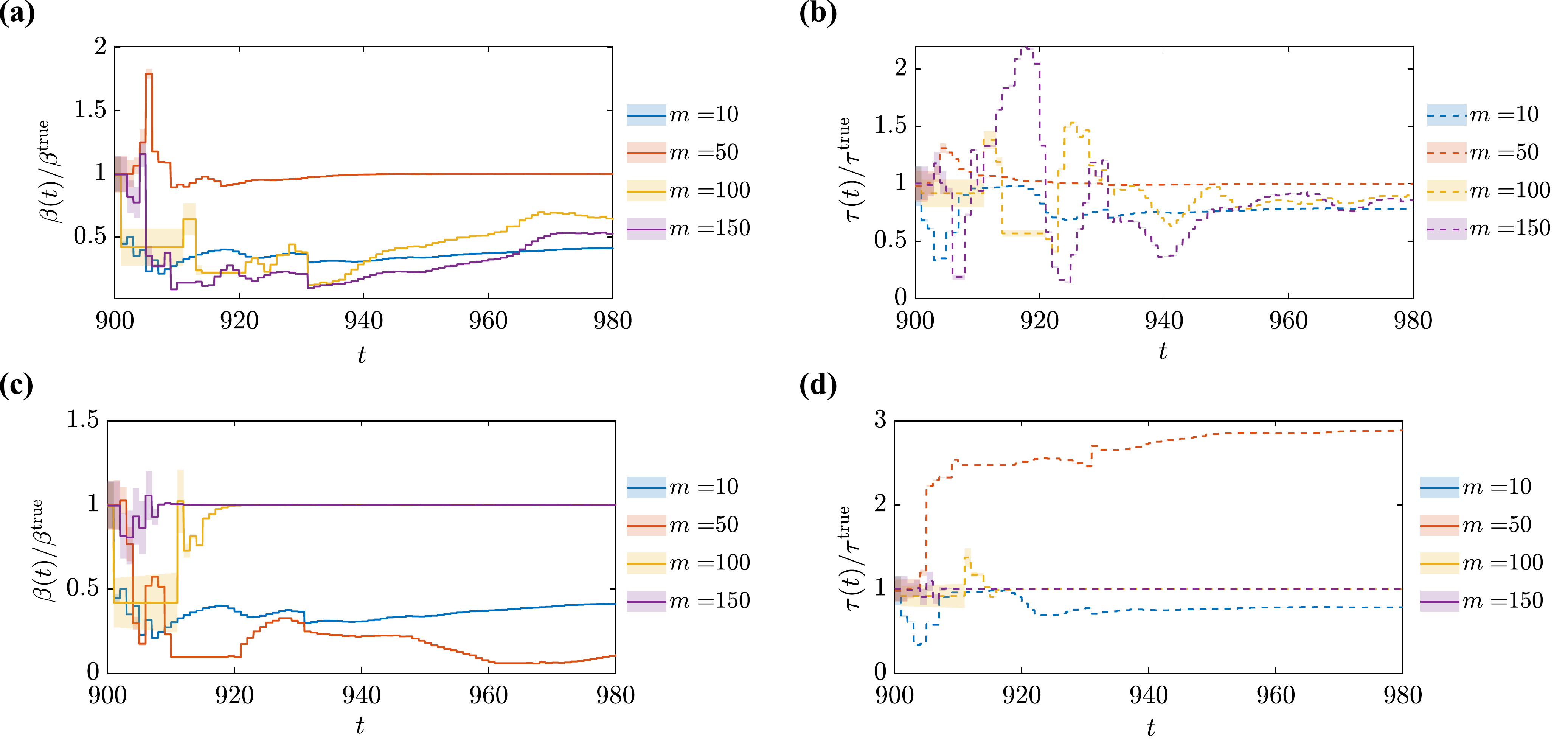}
    \caption{Real--time learning of the parameters. Assimilation of acoustic pressure from microphones for combined state and parameter estimation of a quasiperiodic solution. Left: normalised $\beta$. Right: normalised $\tau$. Effect of ensemble size without inflation (top) and with inflation using $\rho = 1.02$ (bottom). $N_\mtxt{mic}=15$, $\beta^\mtxt{true}=3.6$, $\tau^\mtxt{true}=0.2$,  $\Delta t\mtxt{_{analysis}}=1$, $\sigma_\mathrm{mic}=0.01$. The shaded areas show the standard deviation, which becomes smaller as more data is assimilated.}
    \label{fig:inflation}
\end{figure}
This \textit{increase} and \textit{reject} approach is not always sufficient to reach convergence. Figures~\ref{fig:inflation}a,b, show the same simulation as in Figure~\ref{fig:increase} with more microphones, $N_\mtxt{mic}=15$, and $\Delta t\mtxt{_{analysis}}=1$. In this case, the filtered solution is not converging even for 150 ensemble members, which is caused by covariance collapse. 
To accelerate the convergence and overcome the spurious correlations of finite--seized ensembles~\citep{evensen_data_2009}, 
we introduce a covariance inflation  to the ensemble forecast when the solution of the analysis step provides unfeasible parameters. 
The inflation method can be used to counteract the variance reduction due to the spurious correlations, and force the model to explore more states. 
Here, we include the model uncertainty as stochastic noise by adding an inflation factor $\rho$ to the ensemble forecast  
\begin{equation}
    {A}^\mtxt{f}_{ij} =  \overbar{{A}}_{ij}^\mtxt{f} + \rho\,{\Psi}^\mtxt{f}_{ij}. 
\end{equation}
In this case, $\rho=1.02$ improves the analysis for the quasiperiodic solution. 
If necessary, adaptive strategies can be designed following~\citet{evensen_data_2009}. 
Figure~\ref{fig:inflation}c,d shows the parameters' convergence for the same ensemble sizes as Figures~\ref{fig:inflation}a,b, but with covariance inflation. 
This is sufficient to remove the plateau caused by the divergence of the EnSRKF to unphysical parameters in large ensembles, thereby speeding up the convergence. 

To summarise, we propose an {\it increase, reject, inflate} strategy to learn the nonlinear dynamics and parameters of thermoacoustics.

\section{Twin experiments in chaotic regimes}\label{sec:CH_results}

\begin{figure}
    \centering
    \includegraphics[width=\textwidth]{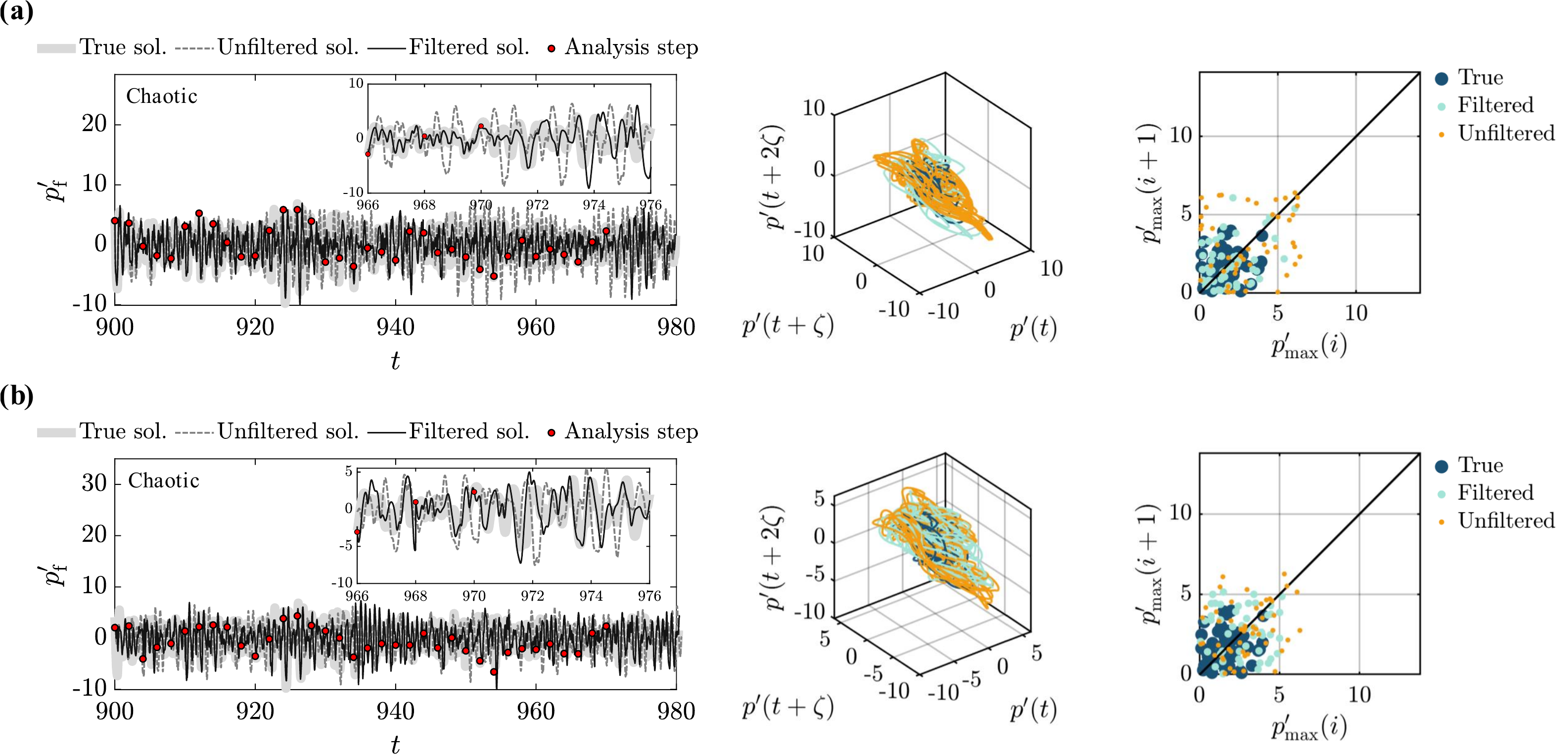}
    \caption{
    {
     Real--time learning of the state. Assimilation of (a) acoustic modes and (b) pressure from microphones for state estimation of a chaotic regime  ($\beta=7.0$). 
  Left: True pressure oscillations at the flame location (light grey), unfiltered solution (dashed dark grey) and  filtered solution (black). The analysis time steps are indicated with red circles. 
  Right: Phase portrait and first return map of the true (dark blue), filtered (light blue), and unfiltered (orange) solutions.
  $m=10$,  $\sigma_\mathrm{mic}=0.01$, $\sigma_\mathrm{frac}=0.25$, $\Delta t\mtxt{_{analysis}}=2$.
    }}
    \label{fig:DA_CH_butterfly}
\end{figure}
\begin{figure}
\centering
    \includegraphics[width=.78\textwidth]{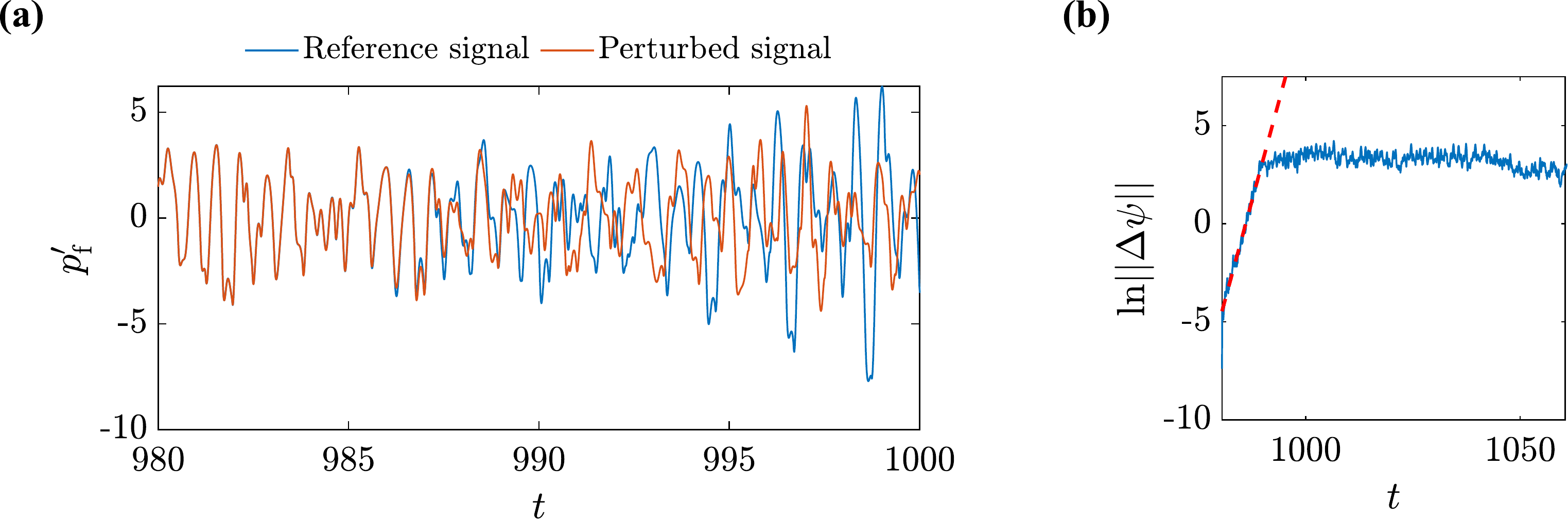}
  \caption{Calculation of the Lyapunov exponent to select the analysis time in data assimilation. (a) Time evolution of the pressure oscillations at the flame location of two nearby chaotic solutions, and (b) logarithmic growth of the trajectory separation.}\label{fig:chaoric_Lyap}
\end{figure}

This section addresses the assimilation in chaotic regimes. 
We perform a series of twin experiments with synthetic data using the base parameters of Tab.~\ref{tab:KF_simulation_parameters} and the obtained suitable parameters in \S~\ref{sec:results}. 
Both, state estimation and combined state and parameter estimation are tested in the chaotic region~F. 
In the combined state and parameter estimation, the initial conditions for $\beta$ and $\tau$ are sampled from uniform distributions with an upper bound 25\% larger than their true value, and a lower bound 25\% smaller than the true parameters.
Different simulations are performed to analyse the predictability of the solutions and to determine a suitable time between analysis ($\Delta t\mtxt{_{analysis}}$), which is not trivial in chaotic oscillations. 

Figure~\ref{fig:DA_CH_butterfly} shows the comparison between the combined state and parameter assimilation solution, an unfiltered solution, and the true state in the chaotic region~F of the bifurcation diagram with the same time between analysis as the previous non--chaotic studies. 
The assimilation does not perform as well as in non--chaotic regimes.
This is physically due to the short predictability of chaotic systems.

\begin{figure}
    \centering
    \includegraphics[width=\textwidth]{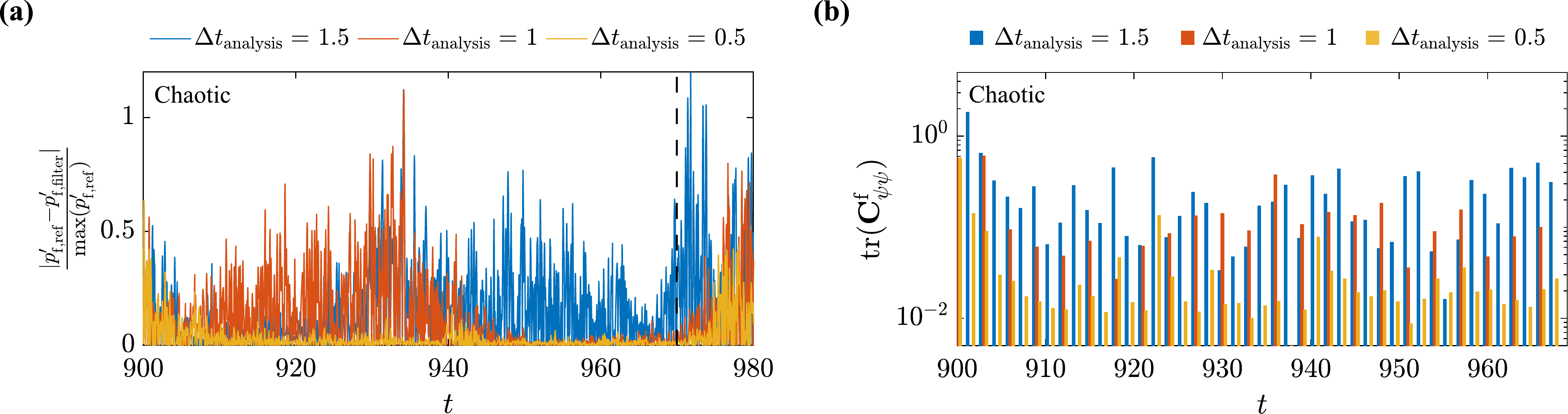}
    \caption{
     Assimilation of acoustic modes for state estimation of a chaotic regime. Performance metrics.  Effect of the assimilation frequency.
    Left: relative difference between the filtered solutions and truth.
    Right: the trace of the ensemble covariance. The dashed vertical line indicates when data assimilation ends. $\beta=7.0$, $m=10$, $\sigma_\mathrm{frac}=0.25$.
    }
    \label{fig:CH_kmeas_error}
\end{figure}

There are several ways to estimate the predictability of a chaotic system~\citep{boffetta_predictability_2002}. 
Here, the predictability is computed as the inverse of the maximal Lyapunov exponent, which provides a time scale after which two nearby trajectories diverge (linearly) due to the butterfly effect. 
The methodology followed is described in~\citet{magri_doan_2019}.  
The maximal Lyapunov exponent is determined by analysing the growth of the distance between two nearby trajectories.  
In a logarithmic scale, the Lyapunov exponent is the slope of the linear region, which is computed by linear regression. 
Figure~\ref{fig:chaoric_Lyap}{a} shows two trajectories that are the same until $t_1= 980$, when they are set apart by $\epsilon = 10^{-6}$. 
After 10 time units, the two instantaneous solutions are completely different, which is a manifestation of chaos. The logarithmic evolution of the distance between the two trajectories is shown in Figure~\ref{fig:chaoric_Lyap}{b}, where the slope of the linear region gives the dominant Lyapunov exponent. 
This method is carried out for several initial conditions in the attractor. The resulting maximal Lyapunov exponent is $\lambda_1=0.74\pm 0.30$, which corresponds to a predictability time scale of $t_\lambda=\lambda_1^{-1}=1.62\pm 0.78$.  
Physically, the predictability, $t_\lambda$, is key to the implementation of the ensemble square--root Kalman filter for time--accurate predictions because, if the time between analysis is too large, the forecast ensemble will already be far apart from the truth. Figure~\ref{fig:DA_CH_butterfly} shows how the filtered chaotic solution with an assimilation time on the high end of the time scale $t_\lambda$ is completely different to the true solution. 
Figure~\ref{fig:CH_kmeas_error} shows the effect of the time between analysis $\Delta t\mtxt{_{analysis}}$ in the chaotic assimilation. 
The EnSRKF time--accurately learns the true solution for $\Delta t\mtxt{_{analysis}}<t_\lambda$ only as the relative error and the trace of the covariance are reduced significantly and converge. 
Therefore, we consider a time between analysis of $\Delta t\mtxt{_{analysis}}=0.5$ for chaotic regions.  
(The butterfly effect is not present in non--chaotic behaviours, therefore, the time  between analysis considered in the fixed point, limit cycle, frequency--locked and quasiperiodic cases can be increased to reduce the computation time, as long as the Nyquist--Shannon criterion is fulfilled~\citep{traverso_data_2019}.) 

\begin{figure}
    \centering
    \includegraphics[width=\textwidth]{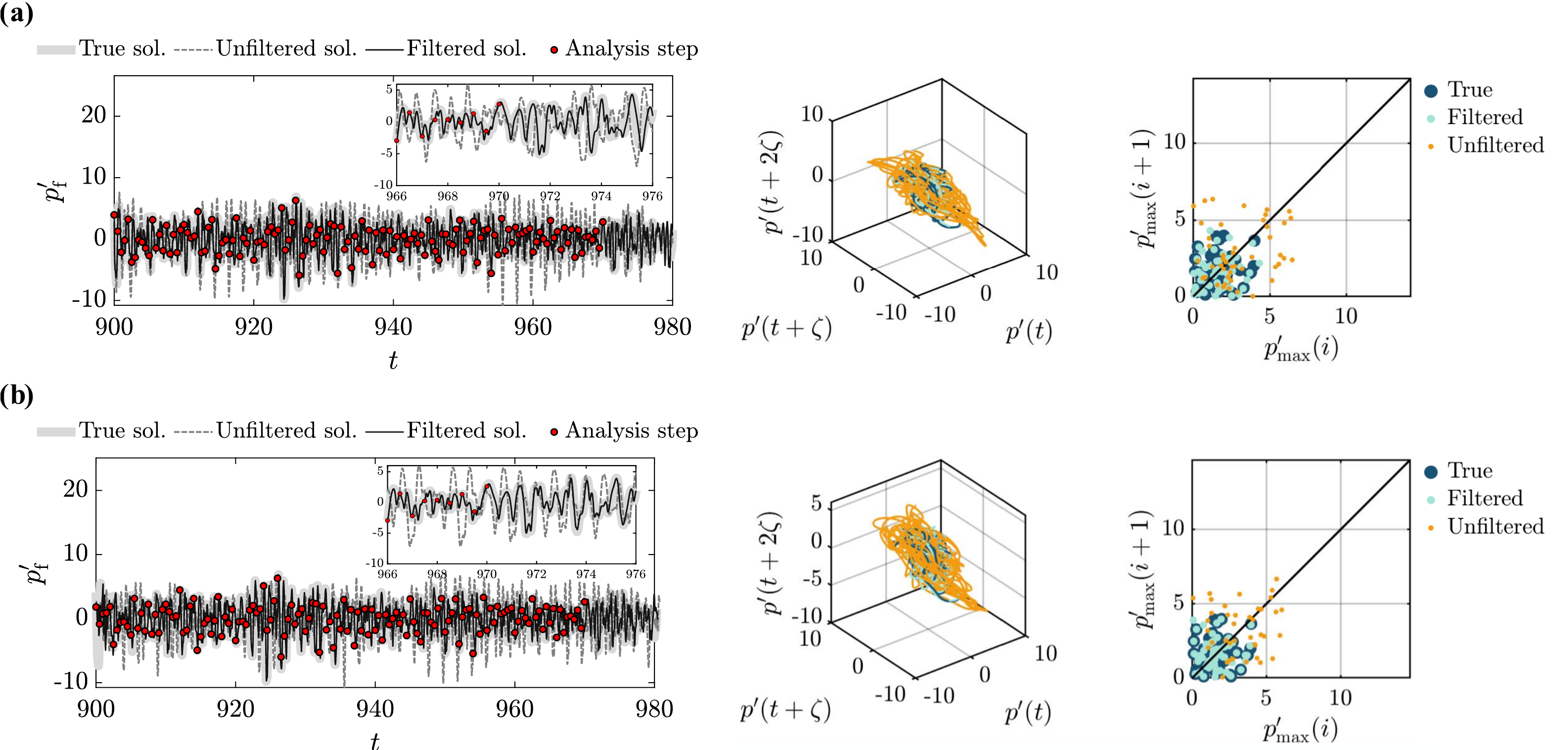}
    \caption{
     {
     Real--time learning of the state. Assimilation of (a) acoustic modes and (b) pressure from microphones for state estimation of a chaotic regime  ($\beta=7.0$). 
  Left: True pressure oscillations at the flame location (light grey), unfiltered solution (dashed dark grey) and  filtered solution (black). The analysis time steps are indicated with red circles. 
  Right: Phase portrait and first return map of the true (dark blue), filtered (light blue), and unfiltered (orange) solutions.
  $m=100$,  $\sigma_\mathrm{mic}=0.01$, $\sigma_\mathrm{frac}=0.25$, $\Delta t\mtxt{_{analysis}}=0.5$.
    }}
    \label{fig:CH_timeseries}
\end{figure}

\begin{figure}
    \centering
    \includegraphics[width=.98\textwidth]{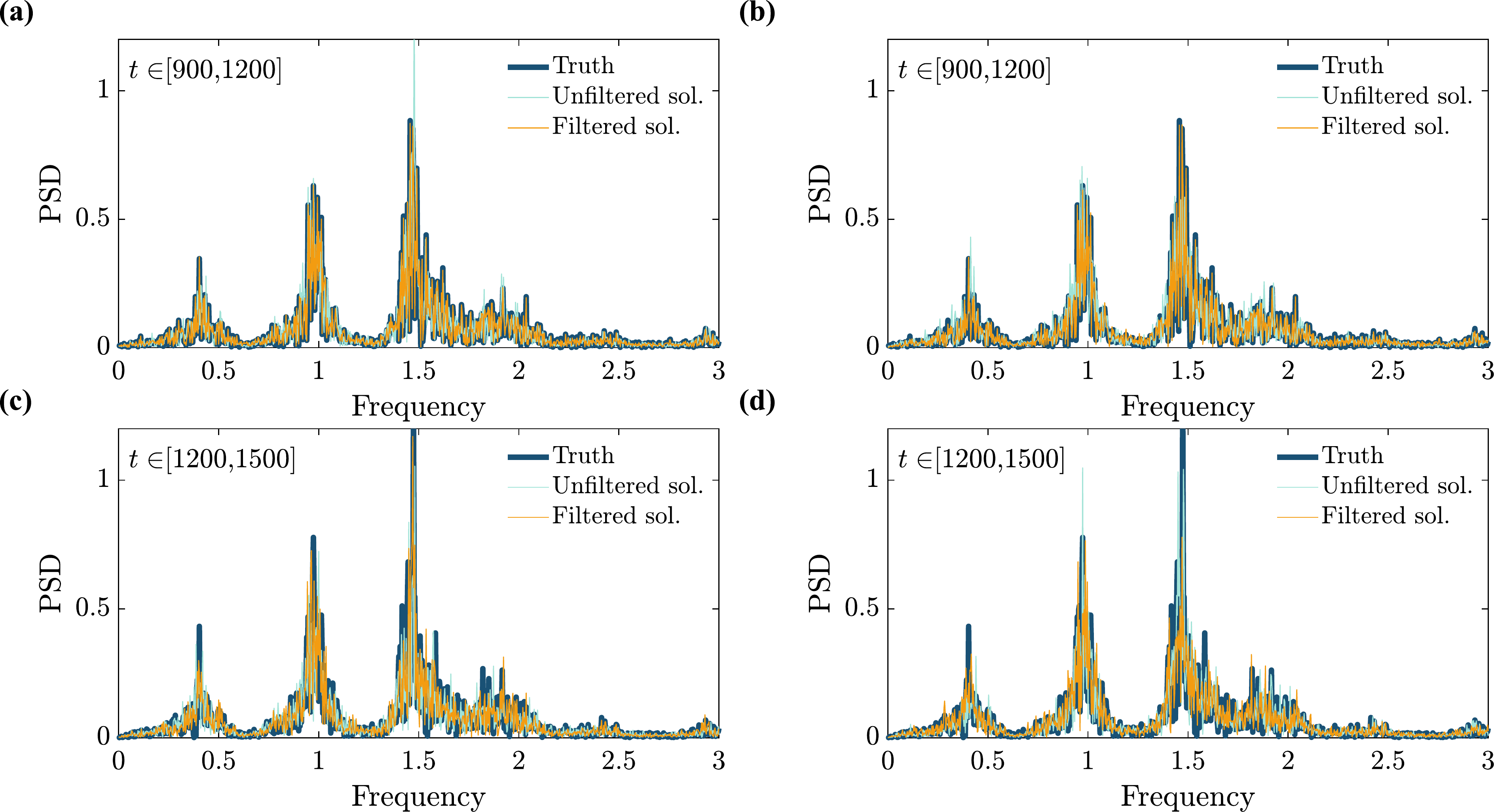}
    \caption{
    Power spectral density (PSD) during (top) and after  (bottom) assimilation of the true pressure oscillations at the flame location (dark blue),  unfiltered solution (light blue) and  filtered solution (orange), during state estimation in a chaotic regime ($\beta=7.0$). The analysis time steps are indicated with red circles. 
    Left: assimilation of acoustic modes. Right: assimilation of pressure from microphones.  
    $m=100$,  $\sigma_\mathrm{mic}=0.01$, $\sigma_\mathrm{frac}=0.25$, $\Delta t\mtxt{_{analysis}}=0.5$.
    }
    \label{fig:CH_PSD}
\end{figure}

Figure~\ref{fig:CH_timeseries} shows the results of state estimation in a chaotic regime. 
The assimilation of the acoustic modes is shown in Figs.~\ref{fig:CH_timeseries}{a}, 
while the assimilation of pressure observations is shown in Figs.~\ref{fig:CH_timeseries}{b}. 
The results are generated with an ensemble of $m=100$. 
The results indicate that the filter learns the pressure state in chaotic regimes for the two assimilation approaches. 
Because of the butterfly effect, the filtered pressure and the true signal start differing after removing the filtering due to the chaotic nature of the solutions. {The agreement is also evident in the phase space reconstruction and first return map.} 
Figure~\ref{fig:CH_PSD} shows the results of state estimation in the form of power spectral density (PSD). 
The top PSDs are computed during the assimilation window ($t\in [900, 1200]$)  and the bottom PSDs are computed after removing the filter and propagating the filtered solution without data assimilation ($t\in [1200, 1500]$). 
The PSDs during the assimilation indicate that the filter learns as well almost exactly the frequency spectrum of the solution, while the unfiltered solution exhibits significant discrepancies. 
After removing the filter, the PSD of the true and filtered solutions remain qualitatively similar, but differ slightly due to the chaotic divergence of the solution. 
\begin{figure}
    \centering
    \includegraphics[width = .9\textwidth]{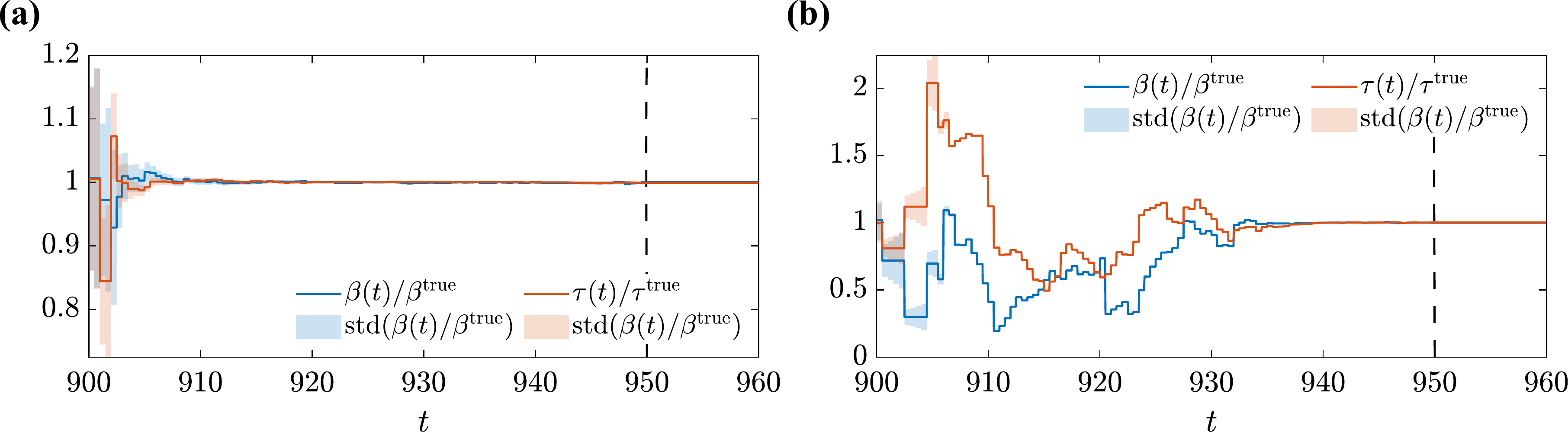}
    \caption{
    Real--time learning of the parameters. Assimilation of (a) acoustic modes and (b) pressure from microphones for combined state and parameter estimation of a chaotic regime. 
    Time evolution of the parameters and their standard deviation.  
    Chaotic solution ($\beta=7.0$). The dashed vertical line indicates when data assimilation end. $m=300$, $\rho=1.2$, $\sigma_\mathrm{mic}=0.01$, $\Delta t\mtxt{_{analysis}}=0.5$, $N_\mtxt{mic} = 6$. %
    }
    \label{fig:Chaos_params_6mic}
\end{figure}
\begin{figure}
    \centering
    \includegraphics[width = .9\textwidth]{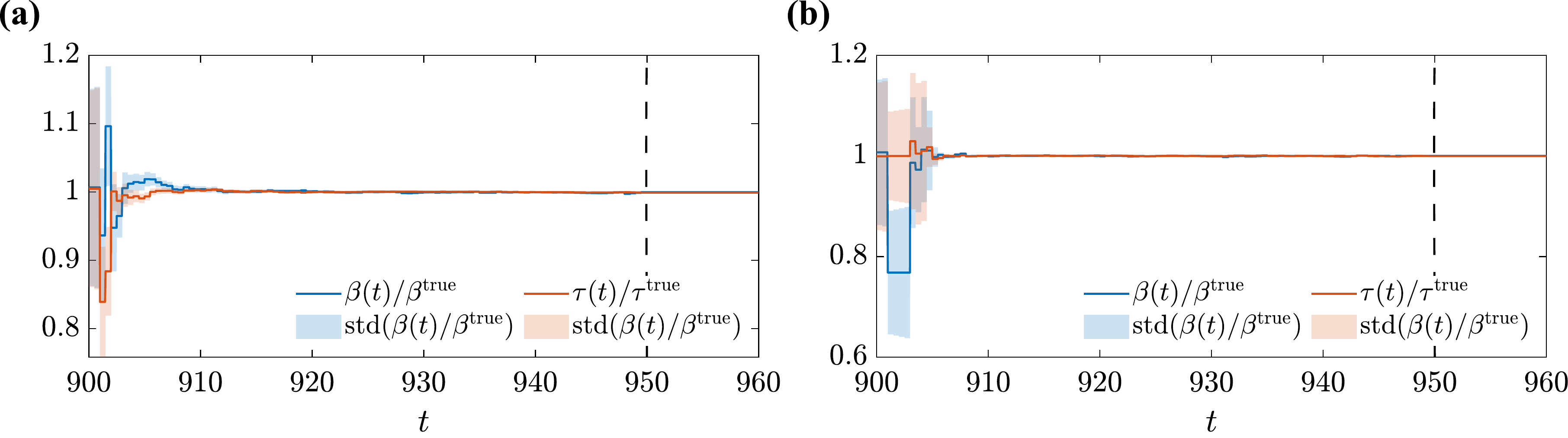}
    \caption{
    Real--time learning of the parameters. Assimilation of (a) acoustic modes and (b) pressure from microphones for combined state and parameter estimation of a chaotic regime. 
    Time evolution of the parameters and their standard deviation. 
    Chaotic solution ($\beta=7.0$). The dashed vertical line indicates when data assimilation ends. $m=300$, $\rho=1.02$, $\sigma_\mathrm{mic}=0.01$, $\Delta t\mtxt{_{analysis}}=0.5$, $N_\mtxt{mic} = 15$. %
    }
    \label{fig:Chaos_params_15mic}
\end{figure}

Finally, the data assimilation algorithm is able to estimate $\beta$ and $\tau$ in the combined state and parameter estimation in chaotic regimes for the assimilation of both acoustic modes and pressure from microphones (Figs.~\ref{fig:Chaos_params_6mic}a,b, respectively). The results indicate that there is a successful convergence of the parameters even though their initial uncertainty is large.  
These simulations are performed with a large ensemble of 300 members and by inflating the ensemble when the assimilation is neglected due to unphysical parameters. The inflation parameter required for convergence in the assimilation of pressure data (Figure~\ref{fig:Chaos_params_6mic}b) is large ($\rho=1.2$). Figures~\ref{fig:Chaos_params_15mic}b shows that the convergence is significantly faster and requires a smaller inflation ($\rho=1.02$) if the number of microphones is increased to 15, as they provide a greater amount of information on the system, i.e., the problem is less ill--conditioned.

The data assimilation successfully learns the true state and parameters for chaotic regimes in the twin experiments by increasing the assimilation frequency, the ensemble size and the inflation parameter. 
%

{
\section{Bias--aware data assimilation with echo state networks}\label{sec:bias_results}
%
The echo state network is trained with the bias, which is the difference between the {statistically stationary} solution of the travelling--wave model with the flame kinematic model~\S~\ref{sec:test_case} and a {statistically stationary} realisation of the lower--fidelity model with initial guess $\tilde{\beta}_\mtxt{train} = 10^6~\mtxt{\frac{W s^{1/2}}{m^{5/2}}}$. The training set is short ($1.2~\mtxt{s}$), and sampled at $10000~\mtxt{Hz}$. This sampling frequency is consistent with experimental works~\citep[e.g.][]{garita_assimilation_2021}. The  washout consists of  $0.025~\mtxt{s}$ of observations sampled at  the same frequency. 
Following the results from~\S~\ref{sec:results}, the simulations in this section use an ensemble size $m=100$, an inflation factor of $\rho = 1.02$, a microphone standard deviation of $\sigma_\mtxt{mic}=0.01$ and a time between analysis of $\Delta\tilde{t}_\mtxt{analysis}=3\E{-3}~\mtxt{s}$.  

\begin{figure}
    \centering
    \includegraphics[width=\textwidth]{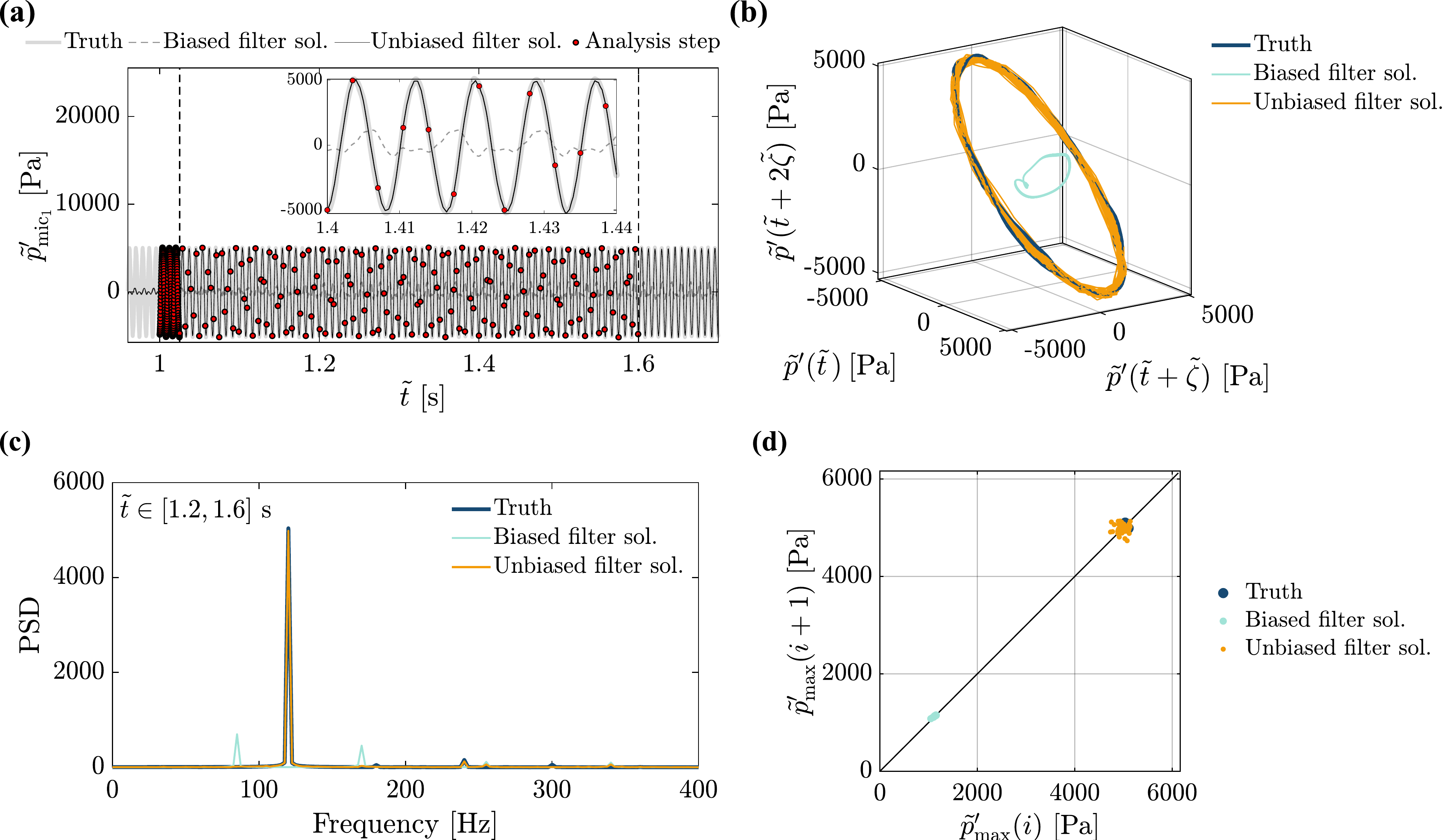}
    \caption{Real--time learning of the model bias and unbiased state. (a) True pressure oscillations at $\tilde{x}/\tilde{L}_0 = 0.18$ (light grey), biased filtered solution (dashed dark grey), and unbiased filtered solution (black). The analysis time steps are indicated with red circles; the vertical dashed lines show the assimilation window. (b) Phase portrait, (c) power spectral density (PSD) and (d) first return map of the true pressure oscillations (dark blue),  biased filtered solution (light blue), and unbiased filtered solution (orange). 
    }
    \label{fig:bias_SE}
\end{figure}
\begin{figure}
    \centering
    \includegraphics[width=\textwidth]{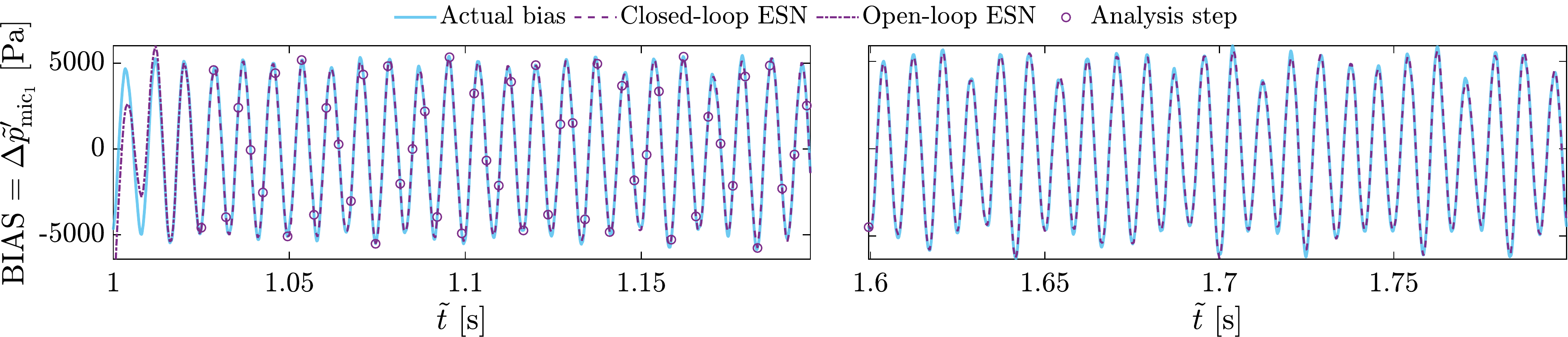}
    \caption{Real--time learning of the model bias and unbiased state. Comparison of the actual bias and the estimated model bias from the echo state network during state estimation. Left: open loop initialisation and sequential closed--loops reinitialised at analysis steps. Right: echo state network's closed--loop forecast after data assimilation ends.}
    \label{fig:bias_SE_ESN}
\end{figure}
Figure~\ref{fig:bias_SE} shows the unbiased state estimation results.  
The biased filtered solution  represents the expectation of the forecast pressure, while the unbiased filtered solution is the resulting pressure state after correcting the model bias with the ESN. 
After~1~second of assimilation--free forecasting, when both the truth and the low--order solutions are {statistically stationary}, the ESN is initialised with  $0.025~\mtxt{s}$ of washout. At $\tilde{t}=1.025~\mtxt{s}$ the state estimation begins. The results indicate that the ESN favourably  estimates the model bias. This allows the EnSRKF to recover the true pressure timeseries, as well as to learn its frequency spectrum and the attractor~(Figure~\ref{fig:bias_SE}{b}). 
%
Figure~\ref{fig:bias_SE_ESN} shows the performance of the ESN at the start of unbiased state estimation, and after the data assimilation ends. 
After the open--loop initialisation, the agreement between the estimated bias and the actual bias is favourable. 
In the autonomous evolution (closed--loop), a re--initialisation every $3\E{-3}~\mtxt{s}$ is sufficient to maintain the accuracy on the inferred bias. 
The ESN is trained with the bias resulting from a simulation using the 
$\tilde{\beta}= \tilde{\beta}_\mtxt{train}$, 
so it is expected to provide good estimates of the bias when initialising the ensemble to the same value of $\tilde{\beta}$, provided a long enough washout.

\begin{figure}
    \centering
    \includegraphics[width=\textwidth]{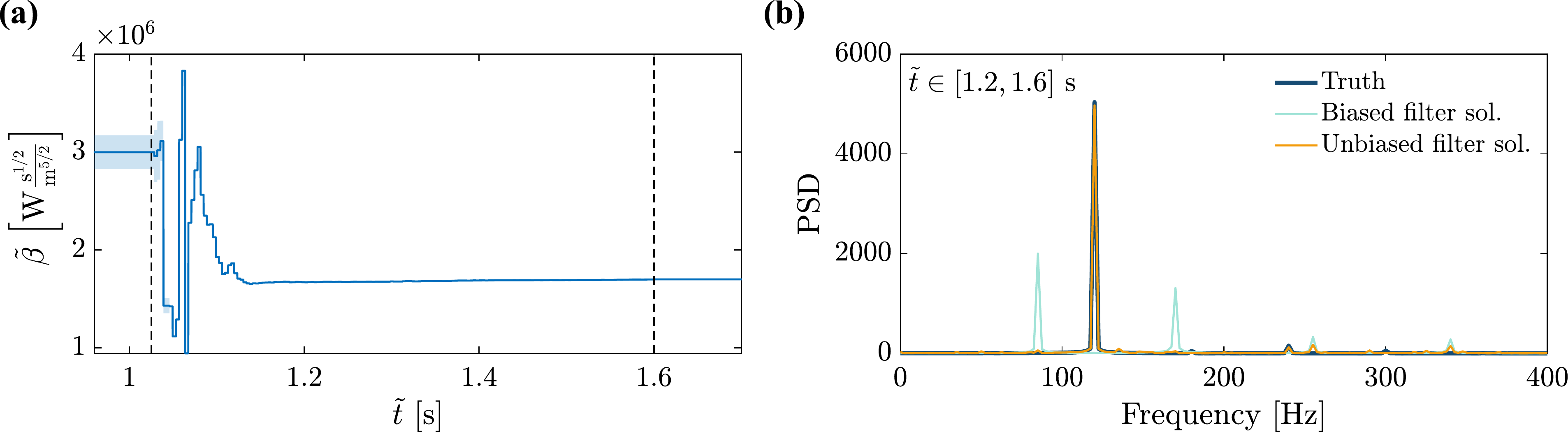}
    \caption{Real--time learning of the model bias, unbiased state and parameters. (a) Time evolution of the parameter $\tilde{\beta}$ with the standard deviation; and (b) power spectral density (PSD) of the pressure oscillations at $\tilde{x}/\tilde{L}_0 = 0.18$ of the true pressure oscillations (dark blue),  biased filtered solution (light blue), and unbiased filtered solution (orange). The vertical dashed lines shows the data assimilation window 
    }
    \label{fig:bias_SPE}
\end{figure}
\begin{figure}
    \centering
    \includegraphics[width=\textwidth]{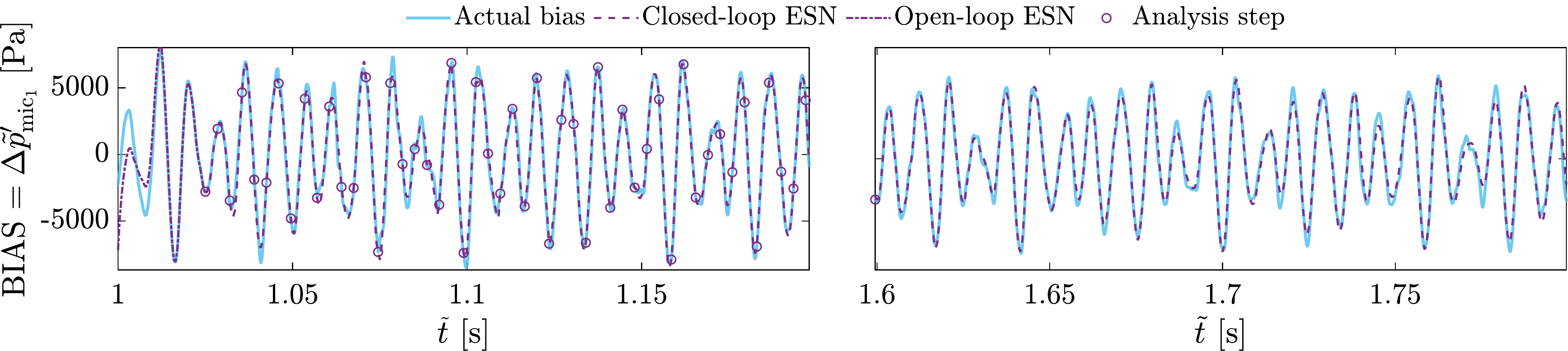}
    \caption{Real--time learning of the model bias, unbiased state and parameters. Comparison of the actual bias and the estimated model bias from the echo state network during state and parameter estimation. Left: open loop initialisation and sequential closed--loops reinitialised at analysis steps. Right: echo state network's closed-loop forecast after data assimilation ends.}
    \label{fig:bias_SPE_ESN}
\end{figure}
The results for combined unbiased state and parameter estimation are shown in Figure~\ref{fig:bias_SPE}. The values of heat source strength for the ensemble members are initialised far from the training $\tilde{\beta}_\mtxt{train}$, as a uniform random distribution with a 10\% standard deviation and a mean value of  $\tilde{\beta}=3\,\tilde{\beta}_\mtxt{train}$. 
The data assimilation with echo state network algorithm converges to a physical value of heat source strength ($\tilde{\beta}=1.70\E{6}~\mtxt{\frac{W s^{1/2}}{m^{5/2}}}$ in Figure~\ref{fig:bias_SPE}{a}), which recovers the dominant frequencies of the higher--fidelity simulation (Figure~\ref{fig:bias_SPE}{b}). 
By comparing Figures~\ref{fig:bias_SPE_ESN} and~\ref{fig:bias_SE_ESN}, it can be seen that the timeseries of the model bias for a low--order model with  $\tilde{\beta}=1.70\E{6}~\mtxt{\frac{W s^{1/2}}{m^{5/2}}}$ is significantly different from that of  $\tilde{\beta}_\mtxt{train}$. 
This means that, although the echo state network is trained on data with a fixed $\tilde{\beta}_{\rm train}=10^{6}~\mtxt{\frac{W s^{1/2}}{m^{5/2}}}$, it is able to infer adaptively the bias of unseen data (with a different $\tilde{\beta}$).  
}

\section{Conclusions}\label{sec:conclusions}

Low--order thermoacoustic models are qualitatively correct, but they are quantitatively incorrect. 
In this work, we introduce data assimilation to make qualitative models quantitatively (more) accurate.
This is achieved by combining the knowledge from observations, such as experimental data, and a physical model prediction. 
Data and model predictions are combined with a Bayesian data assimilation. 
The algorithm learns the state, such as the acoustic pressure, and the parameters of the model, every time that reference data becomes available (real--time). 

%

%
First, we develop a sequential data assimilation algorithm based on the ensemble square--root Kalman filter in the time domain. 
This nonlinear filter selects the most likely state and set of physical parameters, which are compatible with model predictions and their uncertainties, and with observations and their uncertainties. 
The filter is physical, i.e., it is not a purely machine learning technique, as it provides estimates that are compatible with the conservation laws, which makes it robust and principled. 
The data, once assimilated, does not need to be stored. 
For the data assimilation, which is based on a Markov assumption, we transform the time--delayed dynamics (non--Markovian) into an initial value problem (Markovian).  

Second, 
we perform twin experiments in each region of the bifurcation diagram with reference data on
(i) the acoustic Galerkin modes, and 
(ii) the acoustic pressure taken from multiple microphones.
On the one hand, in non--chaotic oscillations, the frequency with which data should be assimilated needs to fulfil the Nyquist--Shannon criterion with respect to the dominant acoustic mode. 
On the other hand, in chaotic oscillations, we highlight that the assimilation frequency should scale with the Lyapunov exponent. 
During the combined state and parameter estimation with pressure observations, it is observed that the filter occasionally provides unphysical solutions, such as negative time delays, which lead to convergence to incorrect solutions. 
This is due to the bifurcations and hystereses that occur in a small range of parameters.
Hence, we propose an {\it increase, reject, inflate} strategy to overcome this.
In detail, we increase the ensemble size to better capture the correct dynamics; we reject the analysis steps that provide unphysical  parameters, e.g., negative time delays; and we inflate the ensemble covariance by adding noise as a regularisation term. 
With the twin experiments in data assimilation, we show that 
(i) the correct acoustic pressure and parameters can be accurately learnt (i.e., inferred); 
(ii) the ensemble size is small (in contrast to standard Kalman filters), from ten to hundred depending on the multi--frequency content of the solution; 
(iii) the learning is robust because it can tackle large uncertainties in the observations (up to 50\% the mean values);   
(iv) the uncertainty of the prediction and parameters is naturally part of the output; and 
(v) both the time--accurate solution and statistics (through power spectral density function) can be successfully learnt. 

{
Third,
we propose a data assimilation framework to learn the model error (bias). 
The model bias is inferred by an echo state network, which is a data-driven tool that is more general than an auto--regressive model. 
We perform data assimilation using reference data from a higher--fidelity acoustic model, which contains a mean flow, non--ideal boundary conditions, and a kinematic model for the flame. The echo state network is trained a priori, and then it is run in parallel with the sequential data assimilation algorithm. 
We show that, with a short training set, the reservoir  learns the dynamics of the thermoacoustic model error. The proposed methodology successfully learns in real time both the time--accurate solution and the statistics of it.}

The technology developed in this paper is being applied to improve the quantitative accuracy of reduced--order models with experimental data from pressure sensors, to learn different model parameters, and to provide estimates of the model error.
Data assimilation with an echo state network opens up new opportunities for real--time prediction of thermoacoustics by synergistically combining physical knowledge and data, as well as for estimating the model bias beyond the field of thermoacoustics.
\vspace{1em}\\
\textbf{Funding}{ A. N. is financially supported by Rolls--Royce, the EPSRC--DTP and the Cambridge Commonwealth, European \& International Trust under a Cambridge European Scholarship. L. M. gratefully acknowledges support from the RAEng Research Fellowships Scheme and the ERC Starting Grant PhyCo  949388. }
\\
\textbf{Declaration of interests}{The authors report no conflict of interest.}
\\
\textbf{Acknowledgements}{
The authors are grateful to Francisco Huhn, who helped the authors produce Figure~\ref{fig:2D_bif_diagram}, and to Alberto Racca, who helped implement the echo state network algorithm. 
}




\appendix

\section{Derivation of the EnSRKF}\label{app:derivation}
Before starting with the derivation of the filter, some definitions are introduced. For  $m$  ensemble members and a state vector $\vect{\psi}_i \in \mathbb{R}^{N \times 1}$, the matrix that encapsulates the ensemble members and the ensemble mean are defined as
\begin{equation}
    \vect{A} = \left(\vect{\psi}_1,\, \vect{\psi}_2,\, \dots,\, \vect{\psi}_m\right)\in \mathbb{R}^{N \times m }\qquad\mtxt{and}\qquad\overbar{\vect{\psi}} \approx \dfrac{1}{m}\,\sum_{i = 1}^m\vect{\psi}_i
\end{equation}
With these, the following definition for the ensemble perturbation matrix applies
\begin{equation}
    \boldsymbol{\Psi} = \left(\vect{\psi}_1 - \overbar{\vect{\psi}},\, \vect{\psi}_2 - \overbar{\vect{\psi}}, \,\dots,\, \vect{\psi}_m-\overbar{\vect{\psi}}\right)
\end{equation}
The ensemble covariance matrix can be determined from \eqref{eq:C_pp_ens}, introducing a factor $(m-1)$ to avoid a sample bias. The covariance matrix is defined as an approximation because it is derived from a statistical sample
\begin{equation}
    \label{eq:C_pp_ens}
    \vect{C}_{{\psi}{\psi}}\approx\dfrac{1}{m-1}\, \boldsymbol{\Psi}\, \boldsymbol{\Psi}^\mtxt{T}
\end{equation}
Accounting for these definitions, the Kalman Filter update \eqref{eq:KF_psia} for the ensembles is in matrix form:
\begin{equation}
\label{eq:Aa_1}
    \vect{A}^\mtxt{a} = \vect{A}^\mtxt{f} + \left(\vect{M}\,\vect{C}^\mtxt{f}_{\psi\psi}\right)^\mtxt{T}
                    \left[\vect{C}_{\epsilon\epsilon} + \vect{M}\,\vect{C}^\mtxt{f}_{\psi\psi}\,\vect{M}^\mtxt{T}\right]^{-1}\left(\vect{Y}-\vect{M}\,\vect{A}^\mtxt{f}\right) 
\end{equation}
where $\vect{Y}\in \mathbb{R}^{q \times m}$ is the matrix containing the $q$ observations of each member in the ensemble; $\vect{M}\in\mathbb{R}^{q \times N}$ is the measurement operator matrix; and  $\vect{C}_{\epsilon\epsilon}\in\mathbb{R}^{q\times q}$ is the observations' error covariance matrix.

Using the definitions for the ensemble covariance in \eqref{eq:C_pp_ens}, the ensemble mean of \eqref{eq:Aa_1} is:
\begin{equation}
\label{eq:Aa_bar}
    \overbar{\vect{A}}^\mtxt{a} = \overbar{\vect{A}}^\mtxt{f} + \vect{\Psi}^\mtxt{f}\,\left( {\vect{M}\,\vect{\Psi}^\mtxt{f}}\right)^\mtxt{T}
                    \left[(m-1)\,\vect{C}_{\epsilon\epsilon} +  {\vect{M}\,\vect{\Psi}^\mtxt{f}}\,\left( {\vect{M}\,\vect{\Psi}^\mtxt{f}}\right)^\mtxt{T}\right]^{-1}\left(\vect{Y}-\vect{M}\,\overbar{\vect{A}}^\mtxt{f}\right)
\end{equation}
where $\overbar{\vect{A}}$ is a $N \times m$ matrix of identical mean analysis states in each column. Introducing now the covariance expression into the analysis error update (see \eqref{eq:KF_C}), yields the analysis covariance matrix
\begin{equation}
\label{eq:C_1}
        \vect{C}^\mtxt{a}_{\psi\psi} = \dfrac{\vect{\Psi}^\mtxt{f}\, {\vect{\Psi}^\mtxt{f}}^\mtxt{T}}{m-1} -
                    \left(\vect{M}\,\dfrac{\vect{\Psi}^\mtxt{f}\, {\vect{\Psi}^\mtxt{f}}^\mtxt{T}}{m-1}\right)^\mtxt{T}
                    \left[\vect{C}_{\epsilon\epsilon} + \vect{M}\,\dfrac{\vect{\Psi}^\mtxt{f}\, {\vect{\Psi}^\mtxt{f}}^\mtxt{T}}{m-1}\,\vect{M}^\mtxt{T}\right]^{-1}
                    \left(\vect{M}\,\dfrac{\vect{\Psi}^\mtxt{f}\, {\vect{\Psi}^\mtxt{f}}^\mtxt{T}}{m-1}\right)
\end{equation}
Equation~\eqref{eq:Aa_bar} and \eqref{eq:C_1} can be simplified by introducing the following matrices
\begin{equation}
\label{eq:SW}
    \vect{S} = \vect{M}\,\vect{\Psi}^\mtxt{f}\qquad \mtxt{and}\qquad    \vect{W_s} =\vect{S}\vect{S}^\mtxt{T} + (m-1)\,\vect{C}_{\epsilon\epsilon}
\end{equation}
leading to
\begin{align}
\label{eq:Aa_bar_SW}
    \overbar{\vect{A}}^\mtxt{a} \;&=\; \overbar{\vect{A}}^\mtxt{f} + \vect{\Psi}^\mtxt{f}\,\vect{S}^\mtxt{T}\,\vect{W_s}^{-1}\left(\vect{Y}-\vect{M}\,\overbar{\vect{A}}^\mtxt{f}\right)\\
    \label{eq:C_a}
        \vect{C}^\mtxt{a}_{\psi\psi} &= \dfrac{1}{m-1}\, \vect{\Psi}^\mtxt{f}\left(\mathbb{I} - \vect{S}^\mtxt{T}\,\vect{W_s}^{-1}\,\vect{S}\right)\,{\vect{\Psi}^\mtxt{f}}^\mtxt{T}\quad
        \therefore \quad \vect{\Psi}^\mtxt{a}\,{\vect{\Psi}^\mtxt{a}}^\mtxt{T} = \vect{\Psi}^\mtxt{f}\left(\mathbb{I} - \vect{S}^\mtxt{T}\,\vect{W_s}^{-1}\,\vect{S}\right)\,{\vect{\Psi}^\mtxt{f}}^\mtxt{T}
\end{align}

The key idea of the EnSRKF is to find a matrix $\vect{\Psi}^\mtxt{a}$ with the covariance of \eqref{eq:C_a}, which is added to the mean ensemble in \eqref{eq:Aa_bar_SW} to compute the full ensemble. First, the matrix $\vect{W_s}$ defined in \eqref{eq:SW} can be eigen--decomposed such that $ \vect{W_s} = \vect{Z}\,{\Lambda}\,\vect{Z}^\mtxt{T}$ because it is a symmetric square matrix, where $\Lambda$ and $\vect{Z}$ are the matrices of eigenvalues (diagonal) and eigenvectors (orthogonal), respectively. Substituting the eigen--decomposition into definition of the analysis perturbation matrix, \eqref{eq:C_a} is re--written as
\begin{equation}
\label{eq:XX}
   \vect{\Psi}^\mtxt{a}\,{\vect{\Psi}^\mtxt{a}}^\mtxt{T} = \vect{\Psi}^\mtxt{f}\left(\mathbb{I} - \vect{S}^\mtxt{T}\,\vect{Z}\,\Lambda^{-1}\,\vect{Z}\,\vect{S}\right)\,{\vect{\Psi}^\mtxt{f}}^\mtxt{T} =\vect{\Psi}^\mtxt{f}\left(\mathbb{I} - \vect{X}^\mtxt{T}\,\vect{X}\right)\,{\vect{\Psi}^\mtxt{f}}^\mtxt{T}
\end{equation}
where  $\vect{X} = \Lambda^{-1/2}\,\vect{Z}^\mtxt{T}\,\vect{S}$. Similarly to the decomposition of $\vect{W_s}$, the symmetric matrix given by the product $\vect{X}^\mtxt{T}\,\vect{X}$ can be expressed as: $\vect{X}^\mtxt{T}\,\vect{X} = \vect{V}\,\vect{\Sigma}\,\vect{V}^\mtxt{T}$, where  $\vect{V}$ is an orthogonal matrix of eigenvectors and $\vect{\Sigma}$ is a diagonal matrix of eigenvalues. Next, introducing this decomposition into \eqref{eq:XX} yields
\begin{align}
   \vect{\Psi}^\mtxt{a}\,{\vect{\Psi}^\mtxt{a}}^\mtxt{T}  &=\;\vect{\Psi}^\mtxt{f}\left(\mathbb{I} - \vect{V}\,\vect{\Sigma}\,\vect{V}^\mtxt{T}\right)\,{\vect{\Psi}^\mtxt{f}}^\mtxt{T}\nonumber\\[0.3em]
   &= \;\vect{\Psi}^\mtxt{f}\,\vect{V}\left(\mathbb{I}-\,\vect{\Sigma}\right)\,\vect{V}^\mtxt{T}\,{\vect{\Psi}^\mtxt{f}}^\mtxt{T}\nonumber\\[0.3em]
   &= \;\left[\vect{\Psi}^\mtxt{f}\,\vect{V}\left(\mathbb{I}-\,\vect{\Sigma}\right)^{\rfrac{1}{2}}\vect{V}^\mtxt{T}\right]\,\left[\vect{\Psi}^\mtxt{f}\,\vect{V}\left(\mathbb{I}-\,\vect{\Sigma}\right)^{\rfrac{1}{2}}\vect{V}^\mtxt{T}\right]^\mtxt{T}
\end{align}
Hence, a solution for the analysis ensemble perturbations, which preserves the zero mean in the updated perturbations and keeps the EnSRKF unbiased, is~\citep{sakov_implications_2008}:
\begin{equation}
    \vect{\Psi}^\mtxt{a} = \vect{\Psi}^\mtxt{f}\,\vect{V}\left(\mathbb{I}-\,\vect{\Sigma}\right)^{\rfrac{1}{2}}\vect{V}^\mtxt{T}
\end{equation}

Finally, the analysis state of the ensembles is determined by adding the analysis ensemble perturbations to the mean analysis ensembles. This analysis state is then propagated in time using the nonlinear forecast model, i.e.:
\begin{subequations}
\begin{align}
    \vect{A}^\mtxt{a} &=  \overbar{\vect{A}}^\mtxt{a} + \vect{\Psi}^\mtxt{a}\\[1em]
    \label{eq:update_a}
    \vect{A}^\mtxt{f}(t+\Delta\,t) &=  \mathcal{F}(\vect{A}^\mtxt{a}(t))
\end{align}
\end{subequations}
where $\mathcal{F}$ is a compact representation of the nonlinear thermoacoustic equations. Note that, in the absence of observations, there would be no data assimilation and the  initial conditions for the next forecast are the forecast states rather than the analysis states, hence
\begin{align}
    \label{eq:update_b}
    \vect{A}^\mtxt{f}(t+\Delta\,t) =  \mathcal{F}(\vect{A}^\mtxt{f}(t))
\end{align}

\end{document}